\documentclass[aps,twocolumn,
amsmath,amssymb]{revtex4-2}
\usepackage{color}
\usepackage{graphicx}
\usepackage{float}
\usepackage{subfigure}
\usepackage{booktabs}
\usepackage{dcolumn}
\usepackage{bm}
\usepackage{amsmath}
\usepackage{accents}
\usepackage{verbatim}
\usepackage{lineno}
\usepackage{tikz,pgfplots}
\usepackage{appendix}
\usepackage{braket}

\newcommand{\beq}{\begin{eqnarray}}
\newcommand{\eeq}{\end{eqnarray}}

\begin{document}
\title{Dynamical theory of the East model. II. A path-probability formalism}

\author{Chong Zha$^{1,2}$}
\author{Yuliang Jin$^{1,2,3}$}
\email{yuliangjin@mail.itp.ac.cn}


\affiliation{$^1$Institute of Theoretical Physics,
 Chinese Academy of Sciences, Beijing 100190, China}
\affiliation{$^2$School of Physical Sciences, University of Chinese Academy of Sciences, Beijing 100049, China}
\affiliation{$^3$Wenzhou Institute, University of Chinese Academy of Sciences, Wenzhou, Zhejiang 325000, China}

\date{\today}

\begin{abstract}
We develop a dynamical theory of the East model---a one-dimensional, hierarchically constrained kinetic Ising model---based on path-probability densities. 
This theory yields an infinite but closed set of equations for single-spin path-probability densities, which, after Laplace transformation, become algebraic difference equations. To render the theory tractable, we introduce a systematic factorization approximation in which the joint probability densities of consecutive flip intervals are factorized into lower-order joint probability densities. We implement the first-order single-interval factorization approximation, obtain a closed set of four equations for the persistence functions, and solve them numerically. The resulting time correlation functions and relaxation times agree closely with Monte Carlo simulations and with the two-domain-basis approximation of the Zwanzig--Mori formalism over a wide range of temperatures, although both approximations fail to capture the exact low-temperature asymptotic super-Arrhenius scaling of the relaxation time. The present formalism provides a plausible foundation for a complete and rigorous theoretical description of the dynamics of the East model and offers a promising route toward generalization to other systems.
\end{abstract}

\maketitle

\section{Introduction}

The theoretical study of glassy dynamics encompasses several approaches developed on the basis of different microscopic models. Two important categories among such models are spin-glass and kinetically constrained models (KCMs). In the former,  quenched disorder is introduced into the Hamiltonian and gives rise to slow dynamics~\cite{parisi1979infinite,parisi1980order}; in the latter,  slow dynamics emerges from imposed kinetic constraints~\cite{fredrickson1984kinetic,jackle1991hierarchically}. Both types of systems can be understood, in a broad sense, in terms of a configuration space with restricted access to a certain portion of it within a given time scale. In spin glasses, such a restricted configuration space originates from a nontrivial energy function with many local minima. In KCMs, by contrast, the energy function is simple, while kinetic constraints block certain paths of dynamical evolution~\cite{chung2001combinatorics}. Systems that would otherwise evolve downhill in the energy landscape may become dynamically trapped because the allowed paths leading to lower-energy configurations are inaccessible. In this sense, kinetic constraints generate an effective rugged landscape in configuration space, even in the absence of a complicated energy function~\cite{zhou2024k, zhou2026spontaneous}. By incorporating kinetic constraints directly into the dynamical rules, KCMs provide a particularly transparent framework for capturing some of the essential dynamical features of structural glasses without explicitly introducing the quenched disorder characteristic of spin glasses~\cite{garrahan2002geometrical, garrahan2003coarse}.

Among the KCMs,  the \textit{East model} is a simple yet representative example. The East model is a one-dimensional Ising chain with an asymmetric kinetic constraint~\cite{jackle1991hierarchically}. The state of the spin at site $i$ is represented by $n_i\in \{0,1\}$, and the system evolves according to Glauber dynamics. The kinetic constraint is directional: a spin can flip only when its right neighbor, $i+1$, is in state $1$; otherwise, it is frozen. When a spin is updated at temperature $T$, it is set to state $1$ with probability $c=1/(1+e^{1/T})$, and to state $0$ with probability $1-c$, where we have set the interaction energy and Boltzmann constant to unity. Consequently, in the low-temperature limit ($T\to 0$, or equivalently $c \to 0$), excitations become increasingly rare. Combined with the kinetic constraint, this leads to a severe reduction in the number of spins that can flip, resulting in extremely slow dynamical relaxation.

Numerical simulations and spectral analyses have established that the East model remains ergodic at any finite temperature, with the only ergodicity-breaking point being the trivial zero-temperature point, $T_{\rm c}=0$ (equivalently, $c^*=0$)~\cite{mauch1999recursive,aldous2002asymmetric}. As the system approaches this point, rigorous spectral results show that the relaxation (correlation) time diverges according to a super-Arrhenius law~\cite{cancrini2008kinetically}
\begin{equation}
    \ln \tau_{\rm corr}= \frac{1}{2T^2\ln{2}}+ o(T^{-2}).
    \label{eq:tau}
\end{equation}

Although the asymptotic low-temperature scaling of the relaxation time has been established, a complete description of the dynamics over the entire time regime is still lacking. For instance, no rigorous theory exists for the spin time-correlation functions at low $T$. An exact solution of this minimal kinetic model is challenging yet fundamentally important. It would play a role analogous to that of the exact solution of the two-dimensional Ising model: the former would shed light on the dynamical slowdown in glass-forming systems, just as the latter has provided foundational insights into thermal critical phenomena~\cite{PhysRev.65.117}.

In the first paper of this series~\cite{Zha2026Dynamical}, we reviewed previous attempts to describe the dynamics of the East model within the Zwanzig--Mori framework, together with the difficulties encountered by these approaches. As discussed there, the infinite hierarchy of equations generated by the Zwanzig--Mori formalism reflects the hierarchical nature of the East-model dynamics~\cite{garrahan2002geometrical}. However, the closure~\cite{wu2004east} and truncation~\cite{van2005glassy} schemes explored so far have neither reproduced the correct asymptotic scaling of the relaxation time nor provided a tractable  route toward an exact description of the dynamics. 
This impasse motivates the present study, in which we propose an alternative theoretical framework that does not rely on the conventional Zwanzig--Mori formalism.


In this paper, we present a theoretical framework  called the \textit{path-probability formalism}. This formalism yields a set of equations for the single-spin path-probability function in equilibrium dynamics. Although the set contains an infinite number of equations, it is closed. This set provides an exact theoretical description of the equilibrium dynamics of the East model, but solving these equations remains out of reach, both analytically and numerically.

Symmetries can provide powerful constraints on dynamical variables associated with different configurations or spatial locations.
In the equilibrium dynamics of the East model, a particularly important symmetry is spatial translational invariance. When the initial configuration is sampled from the equilibrium distribution, the marginal path-probability density of a given spin is identical to that of any other spin, owing to the translational invariance of the system. This translational invariance provides the basis for formulating the dynamics in terms of single-spin marginal path-probability densities.

Building on this symmetry, we characterize the dynamical history of a spin by the sequence of time intervals during which it remains in a given state, and introduce the corresponding path probability densities. Owing to the directional kinetic constraint of the East model, the evolution of a spin depends on the dynamical history of its right-hand environment. By decomposing this environmental dependence in real space and exploiting spatial translational invariance, we can express the resulting relations entirely in terms of the marginal path-probability densities of a single spin. In this way, we establish an exact, formally closed system of equations for these marginal path-probability densities, without introducing any closure approximations.

Although this system is formally closed, it  involves an infinite number of coupled path-probability densities and is therefore not directly tractable. To address this difficulty, we introduce a systematic factorization approximation, in which the joint probability densities of successive spin-flip intervals are factorized into products of lower-order joint probability densities. By retaining correlations within groups of $k$ consecutive flip intervals while neglecting correlations between different groups, the infinite system can be reduced to a finite set of equations at each approximation order. Furthermore, by applying Laplace transforms, the original multiple-integral equations are converted into algebraic difference equations, providing a convenient formulation for analytical and numerical investigations. In this work, we develop and examine the first- and second-order approximations, with particular emphasis on the first-order approximation, i.e., the single-interval factorization (1-IF) approximation. We calculate the time correlation functions and relaxation times, and compare the results with those of the most advanced available approach within the Zwanzig--Mori framework, namely, the two-domain basis (2-DB) approximation theory~\cite{van2005glassy}. 
The close agreement between the 1-IF and 2-DB theories
 motivates us to explore possible connections between their underlying approximation schemes.

The rest of the paper is organized as follows. In Sec.~II, we define the East model, introduce the dynamical descriptors, and describe the simulation methods. In Sec.~III, we present the general framework of the dynamical path-probability approach. In Sec.~IV, we develop the dynamical path-probability theory specific to the East model. In Sec.~V, we reformulate the path-probability equations in the Laplace domain. In Sec.~VI, we introduce the systematic factorization approximation, develop the 1-IF scheme, compare the results with Monte Carlo simulations and with several theories within the Zwanzig--Mori framework, and apply our theory to small systems. Section~VII concludes our results and discusses future directions. Technical details of the Laplace transform, the derivation of the kernel operator, the two-interval factorization (2-IF), and the comparison with the 2-DB approximation are provided in Appendices~A--D, respectively.

\section{Model and simulation methods}
\subsection{Model definition}

The East model is a one-dimensional chain of binary (Ising) spins, \(n_i \in \{0,1\}\), evolving according to Glauber dynamics.
The model has a noninteracting Hamiltonian
\begin{equation}
H=\sum_i n_i,
\end{equation}
with an equilibrium distribution
\begin{equation}
    P_{\mathrm{eq}}(\mathbf{n})=\frac{1}{Z}e^{-\beta\sum_i n_i},
    \label{eq:EquilibriumDistribution}
\end{equation}
where \(\beta = 1/T\) is the inverse temperature, \(Z\) is the partition function, and \(\mathbf{n} = \{n_0, \ldots, n_i, \ldots, n_N\}\).
Because this equilibrium distribution factorizes, the model has no static correlations.

Although the equilibrium distribution is simple, the dynamics is nontrivial.
The kinetic constraint rule permits a spin to flip only if its right neighbor is in the excited state, \(n_{i+1}=1\).
When this constraint is satisfied, the flip rates for the \(0\to1\) and \(1\to0\) transitions are \(c\) and \(1-c\), respectively.
The stochastic dynamics of the East model is therefore governed by the master equation
\begin{equation}
    \partial_t P(\mathbf{n},t)
    =\sum_i\left[\tilde{\omega}_{\bar{n}_i}P(\mathbf{n}',t)-\tilde{\omega}_{n_i}P(\mathbf{n},t)\right]
    = -\mathcal{L}P(\mathbf{n},t),
\label{eq:masterEQ}
\end{equation}
where \(\mathcal{L}\) is the Markov operator, \(\mathbf{n}' = \{n_0, \ldots, \bar{n}_i, \ldots, n_N\}\), and \(\bar{n}_i=1-n_i\).
In this equation,
\begin{equation}
\tilde{\omega}_{n_i}
= \omega_{n_i} n_{i+1}
= [c+(1-2c)n_i]n_{i+1}
\label{eq:tildeomega}
\end{equation}
denotes the transition rate for the process \(n_i\to \bar{n}_i\), and \(P(\mathbf{n},t)\) is the probability of configuration \(\mathbf{n}\) at time $t$.
Note that
\begin{equation}
\omega_{n_i} = c+(1-2c)n_i
\label{eq:omega}
\end{equation}
is the bare transition rate, i.e., the rate without the facilitation constraint.

The detailed balance condition reads:
\begin{equation}
P_{\mathrm{eq}}(n_i)\,\omega_{n_i}
=
P_{\mathrm{eq}}(\bar{n}_i)\,\omega_{\bar{n}_i}.
\end{equation}
Together with the single-spin transition rates, this condition yields
\begin{equation}
c=\frac{1}{1+e^{\beta}}.
\end{equation}
The equilibrium probability that \(n_i=1\) is
\begin{equation}
P_{\mathrm{eq}}(n_i=1)
=
\frac{e^{-\beta}}{1+e^{-\beta}}
=
c.
\end{equation}
Therefore, in an equilibrium configuration, the fractions of up and down spins are \(c\) and \(1-c\), respectively.

\subsection{Time correlation function and return probability}

The relaxation process of the East model in equilibrium is characterized by the auto-correlation function of single spin fluctuations: 
\begin{equation}
    \begin{split}
    C(t)&=\frac{\langle\delta n_i(t)\delta n_i(0)\rangle}{c(1-c)}\\
    &=\frac{P[n_i(t)=1|n_i(0)=1]-c}{1-c}
    ,
    \end{split}
    \label{eq:Paper2_CorrFunc}
\end{equation}
where $\delta n_i=n_i-c$.
We refer to $P[n_i(t)=1| n_i(0)=1]$ as the {\it return probability}: the probability that $n_i$ is up at time $t$, given that it was up at time $0$, with any number (including zero) of intermediate flips permitted.

The corresponding relaxation (correlation) time is defined as
\begin{equation}
    \tau_{\rm corr}=\int_0^{\infty}{C(t)dt}.
\label{eq:Paper2_CorrRelaxationTime}
\end{equation}
The correlation function $C(t)$ and time $\tau_{\rm corr}$ are the key objectives for the dynamical theory of the East model. Note that the cross-correlation between two different spins ($i \neq j$) is always zero, $\langle\delta n_j(t)\delta n_i(0)\rangle = 0$~\cite{pitts2000facilitated, van2005glassy}.

\subsection{Discrete-time Monte-Carlo and continuous-time Markov  dynamics}

The dynamics of the East model can be simulated either by a discrete-time Monte Carlo procedure or  a continuous-time Markov process. The two approaches are equivalent in the large-system limit. For the theoretical derivation below, the continuous-time Markov perspective is more convenient.

{\it (i) Discrete-time Monte Carlo dynamics}. At each Monte Carlo step one randomly selects one spin $n_i$, out of the \(N\) spins, to attempt an update.
If its neighbor $n_{i+1} =1$, this spin is allowed to flip, with the bare flip rate given by \(\omega_{n_i}\) in Eq.~(\ref{eq:omega}). The probability that a given spin is selected at each step is \(1/N\), so the probability that it remains unselected after \(k\) steps is $p_{\mathrm{renewal}}(k)=\left(1-1/N\right)^k$.
Setting \(t=k/N\) and taking the limit \(N\to\infty\), one obtains
\begin{equation}
p_{\mathrm{renewal}}(t)
=
e^{-t}.
\label{eq:p_renewal}
\end{equation}
Here \(t\) is the waiting time until the spin is selected, which we call an \textit{attempt event} or \textit{renewal event}. The exponential distribution of \(t\) shows that renewal events follow a Poisson process of unit rate.

Similarly, the waiting-time distribution for the first \textit{flipping event} is
\begin{equation}
p_{\mathrm{flip}}(t)
= \omega_{n_i} e^{-\omega_{n_i}t}.
\label{eq:p_flip}
\end{equation}
Note that Eq.~(\ref{eq:p_flip}) does not include facilitation effects. These waiting-time distributions will be used in our theory below.

{\it (ii) Continuous-time Markov process}.
When a spin \(n_i\) is facilitated (\(n_{i+1}=1\)), a renewal event occurs with a waiting time drawn from Eq.~(\ref{eq:p_renewal}). At each renewal event, the spin is redrawn from equilibrium: it takes the value \(1\) with probability \(c\) and \(0\) with probability \(1-c\). This renewal process is memoryless and depends only on the current state of the system.
A renewal event does not necessarily change the spin state; the spin state changes only when a flipping event occurs.
 The waiting-time distribution for the first flipping event is the same as that in Eq.~(\ref{eq:p_flip}). This dynamics is rejection-free, and thus more efficient than the Monte Carlo approach.

\begin{figure}
    \centering
    \includegraphics[width=0.9\linewidth]{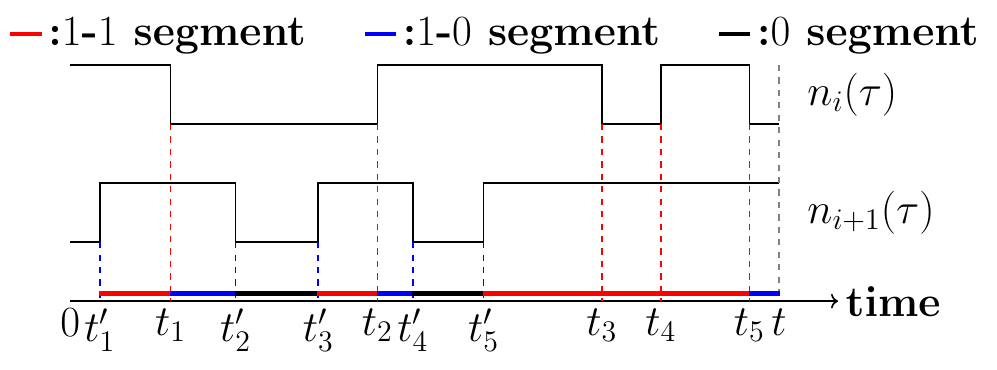}
    \caption{A pair of  trajectories of two neighboring spins, for which the kernel takes the form $\mathcal{Q}^{10}[\{n_i(\tau)\}_t|\{n_{i+1}(\tau)\}_t]=\omega_1\exp[-\omega_1(t_1-t'_1)]\times \exp[-\omega_0(t'_2-t_1)]
        \times\omega_0\exp[-\omega_0(t_2-t'_3)]\times \exp[-\omega_1(t'_4-t_2)]
        \times\omega_1\exp[-\omega_1(t_3-t'_5)]
        \times\omega_0\exp[-\omega_0(t_4-t_3)] \times \omega_1\exp[-\omega_1(t_5-t_4)] \times \exp[-\omega_0(t-t_5)]$.
      }
    \label{fig:Paper2_PathOfTwoSpin}
\end{figure}

\section{General framework of the dynamical path-probability approach}

A primary goal of the dynamical theory is to solve for the time correlation function in Eq.~(\ref{eq:Paper2_CorrFunc}). To this end, it suffices to know the probability $\mathcal{P}_i[\mathcal{C}_i^t(\tau)]$ of every possible trajectory $\mathcal{C}_i^t(\tau) = \{n_i(\tau)\}_{0 \leq \tau \leq t}$, where $\mathcal{C}_i^t(\tau)$ denotes the trajectory of $n_i$ over the interval $0\le \tau \le t$. 
Below we use $\{n_i(\tau)\}_{t}$ as shorthand  for $\{n_i(\tau)\}_{0 \leq \tau \leq t}$.

The single-spin trajectory probability of $n_i$ can be related to that of the neighboring spin $n_{i+1}$ through a conditional probability. Figure~\ref{fig:Paper2_PathOfTwoSpin} shows an example of $\mathcal{C}_i^t(\tau)$ and $\mathcal{C}_{i+1}^t(\tau)$.
This dependence yields a closed equation under equilibrium conditions.
This is the main idea of the path-probability framework. In the following, we describe this framework in a general formalism.

Let \(A\) be the subsystem containing the local variables of interest, and \(B\) the environmental system that influences the dynamics of \(A\). For the East model, \(A\) consists of a single spin \(n_i\), and \(B\) is its right neighbor \(n_{i+1}\). The path-probability density functional \(\mathcal{P}_A[\mathcal{C}^t_A(\tau)]\) of \(A\) is obtained by summing over all possible trajectories of subsystem \(B\),
\begin{equation}
    \mathcal{P}_A[\mathcal{C}^t_A(\tau)]=\sum_{\{\mathcal{C}^t_B(\tau)\}}\mathcal{Q}[\mathcal{C}^t_A(\tau)|\mathcal{C}^t_B(\tau)]\mathcal{P}_B[\mathcal{C}^t_B(\tau)].
    \label{eq:Paper2_ProbAFinal}
\end{equation}
Here, \(\mathcal{Q}[\mathcal{C}^t_A(\tau)|\mathcal{C}^t_B(\tau)]\) is the conditional path-probability density functional, i.e., the probability density that subsystem \(A\) follows the trajectory \(\mathcal{C}^t_A(\tau)\), given that subsystem \(B\) follows the trajectory \(\mathcal{C}^t_B(\tau)\) over the interval \(0\leq \tau \leq t\). For systems with continuous state spaces, the summation over trajectories is replaced by a functional integral.

Equation~(\ref{eq:Paper2_ProbAFinal}) is not closed as it stands: \(\mathcal{P}_A\) for subsystem \(A\) depends on  \(\mathcal{P}_B\) of its environment \(B\), which itself depends on the dynamics \(\mathcal{P}_C\) of \(B\)'s own environment \(C\). However, for the East model the situation is much simpler.  In equilibrium, both \(A=n_i\) and \(B=n_{i+1}\) are described by the same \(\mathcal{P}^{\rm eq}\). Thus Eq.~(\ref{eq:Paper2_ProbAFinal}) becomes closed:
\begin{equation}
    \mathcal{P}^{\rm eq}[\mathcal{C}^t_A(\tau)]=\sum_{\{\mathcal{C}^t_B(\tau)\}}\mathcal{Q}[\mathcal{C}^t_A(\tau)|\mathcal{C}^t_B(\tau)]\mathcal{P}^{\rm eq}[\mathcal{C}^t_B(\tau)].
    \label{eq:Paper2_ProbeqFinal}
\end{equation}

\section{Dynamical path-probability formalism of the east model}

\subsection{A single-spin trajectory as a sequence of flipping events}

Since the East model has discrete spin states, the trajectory of a single spin is completely specified by its initial state and the sequence of its flip times. We therefore characterize a single-spin trajectory by
\begin{equation}
   \{n_i(\tau) \}_{t}
   = \{n_i(0);t_1,t_2,\dots,t_m, t\}
   = \{n_i(0);{\bf t}_m, t\}
    \label{eq:Paper2_TrajectoryExp}
\end{equation}
where \(t_1,\dots,t_m\) are the successive flip times within the observation interval \([0,t]\).

To facilitate the Laplace transform, we also introduce the vector of time intervals between successive state changes,
\[
\Delta {\bf t}_{m+1}=\{t_1,t_2-t_1,\dots,t-t_m\}.
\]
A specific trajectory starting from the initial state \(n_i(0)=n_i\), undergoing \(m\) flips at times \(t_1,t_2,\dots,t_m\), and remaining in its final state until time \(t\), is assigned the path-probability density
\begin{equation}
    \Lambda_m^{n_i}(\Delta {\bf t}_{m+1})
    = \mathcal{P}_i[\{n_i(0);{\bf t}_m, t\}].
    \label{eq:Paper2_PathProbDensityExplicit}
\end{equation}
Here, the notation \(\Lambda_m^{n_i}(\Delta {\bf t}_{m+1})\) shows that the path-probability density for a given trajectory is uniquely specified by the initial state \(n_i = n_i(0)\), the number of flips \(m\), and the time-interval vector \(\Delta {\bf t}_{m+1}\). This parametrization allows the sum over single-spin trajectories to be decomposed into a sum over the number of flips and an integral over the corresponding time intervals:
\begin{equation}
\begin{split}
   \sum_{{\{\{n_i(\tau)\}_{t} \}}}
   &= \sum_{m=0}^{\infty}\int_{0<t_1<\dots<t_m<t}dt_1dt_2\dots dt_m \\
   &= \sum_{m=0}^{\infty}
\int_0^\infty
\left(\prod_{\alpha=1}^{m+1}d\Delta t_\alpha\right)
\delta\!\left(t-\sum_{\alpha=1}^{m+1}\Delta t_\alpha\right)
   ,
   \end{split}
   \label{eq:Paper2_sumToTimeIntegrate}
\end{equation}
where the outer braces in ${\{\{n_i(\tau)\}_{t} \}}$ denote the set of all possible trajectories \(\{n_{i}(\tau)\}_t\).

\subsection{Return probability}

Equation~(\ref{eq:Paper2_CorrFunc}) shows that the correlation function is related to the return probability
\(P[n_i(t)=1| n_i(0)=1]\), a conditional probability. This return probability is obtained by summing the path-probability densities conditioned on $n_i(0)=1$ over all trajectories that terminate in the state $n_i(t)=1$::
\begin{equation}
    \begin{split}
        &P[n_i(t)=1\mid n_i(0)=1] \\
        &=\sum_{\substack{\{\{n_i(\tau)\}_t\}\\n_i(t)=1}}\mathcal{P}_i[\{n_i(\tau)\}_t\mid n_i(0)=1].
    \end{split}
    \label{eq:Paper2_P11SchematicCal}
\end{equation}
which establishes the connection between the correlation function and the path-probability functions via Eq.~(\ref{eq:Paper2_CorrFunc}).
Since both the initial and final states are equal to \(1\), only trajectories with an even number of flips contribute. Combining the notations in Eq.~(\ref{eq:Paper2_PathProbDensityExplicit}) with Eq.~(\ref{eq:Paper2_sumToTimeIntegrate}), we therefore obtain
\begin{equation}
    \begin{split}
        &P[n_i(t)=1| n_i(0)=1] {}\\
        &=\sum_{m=0}^{\infty}
        \int_{0<t_1<\cdots<t_{2m}<t} \left( \prod_{\alpha=1}^{2m} dt_\alpha \right)
        \Lambda_{2m}^1(\Delta {\bf t}_{2m+1}) \\
        & = \sum_{m=0}^{\infty}
        \int_0^\infty \left( \prod_{\alpha=1}^{2m+1} d \Delta t_\alpha \right) \delta \left( t - \sum_{\alpha=1}^{2m+1} \Delta t_\alpha \right)
        \Lambda_{2m}^1(\Delta {\bf t}_{2m+1}).
    \end{split}
    \label{eq:Paper2_CorrToTrajectory}
\end{equation}
Here and throughout the rest of this paper, we adopt the convention that $\prod_{\alpha=1}^0=1$.

\subsection{An infinite closed system of path-probability equations in equilibrium}

In equilibrium, spin \(i\) and its right neighbor $i+1$ have exactly the same marginal path-probability density
\begin{equation}
\mathcal{P}^{\rm eq}[\{n_i(\tau)\}_t]
=
\mathcal{P}^{\rm eq}[\{n_{i+1}(\tau)\}_t].
\end{equation}
Substituting \(A=n_i\) and \(B=n_{i+1}\) into Eq.~\eqref{eq:Paper2_ProbAFinal} yields
\begin{align}
\begin{split}
    &\mathcal{P}^{\rm eq}[\{n_i(\tau)\}_t]={}\\
    &\sum_{\{\{n_{i+1}(\tau)\}_t\}}
    \mathcal{Q}[\{n_i(\tau)\}_t | \{n_{i+1}(\tau)\}_t] 
    \,
    \mathcal{P}^{\rm eq}[\{n_{i+1}(\tau)\}_t].
\end{split}
\label{eq:Paper2_EqProbDecompInEastModel}
\end{align}

For equilibrium dynamics, the initial configuration is sampled from the equilibrium distribution, so that
\(n_i(0)=1\) or \(n_i(0)=0\) with probabilities \(c\) and \(1-c\), respectively. Thus,
\begin{equation}
\begin{split}
    \mathcal{P}^{\rm eq}[\{n_i(\tau)\}_t]
    ={}& c\,\mathcal{P}^{\rm eq}[\{n_i(\tau)\}_t | n_i(0)=1]\\
    &+(1-c)\,\mathcal{P}^{\rm eq}[\{n_i(\tau)\}_t | n_i(0)=0].
\end{split}
\label{eq:Paper2_EquilibriumDecompOfSingleSpin}
\end{equation}
Here,  \(n_i(0)=1\) or \(n_i(0)=0\) fixes the initial state of spin \(i\), while the initial states  of all other spins are sampled from the equilibrium distribution and summed over. Thus,
\(\mathcal{P}^{\rm eq}[\{n_i(\tau)\}_t | n_i(0)= n_i]\)
denotes the probability of the trajectory \(\{n_i(\tau)\}_t\) in equilibrium dynamics given \(n_i(0)=  n_i\), with all other spins marginalized out.

Using Eq.~\eqref{eq:Paper2_EquilibriumDecompOfSingleSpin}, we can rewrite
Eq.~\eqref{eq:Paper2_EqProbDecompInEastModel} as
\begin{widetext}
\begin{equation}
\begin{split}
    \mathcal{P}^{\rm eq}[\{n_i(\tau)\}_t|n_i(0)=n_i]
    ={}&
    \sum_{\{\{n_{i+1}(\tau)\}_t\}}
    \Big\{
    c\,\mathcal{Q}^{n_i 1}[\{n_i(\tau)\}_t|\{n_{i+1}(\tau)\}_t]\,
    \mathcal{P}^{\rm eq}[\{n_{i+1}(\tau)\}_t|n_{i+1}(0)=1]\\
    &+(1-c)\,\mathcal{Q}^{n_i 0}[\{n_i(\tau)\}_t|\{n_{i+1}(\tau)\}_t]\,
    \mathcal{P}^{\rm eq}[\{n_{i+1}(\tau)\}_t|n_{i+1}(0)=0]
    \Big\},
\end{split}
\label{eq:Paper2_FinalShematicEqSystem}
\end{equation}
\end{widetext}
where \(n_i=0\) or \(1\). The superscript in
\(\mathcal{Q}^{n_i n_{i+1}}\) explicitly indicates the initial conditions
\(n_i(0)=n_i\) and \(n_{i+1}(0)=n_{i+1}\).

Using the notation introduced in Eq.~(\ref{eq:Paper2_PathProbDensityExplicit}), 
we can write Eq.~(\ref{eq:Paper2_FinalShematicEqSystem}) in a more explicit form, 
indicating the number of flips \(m\) and writing out the multi-time integral (see Eq.~\ref{eq:Paper2_sumToTimeIntegrate}):
\begin{widetext}

\begin{equation}
\begin{aligned}
\Lambda_m^{n_i}(\Delta\mathbf t_{m+1})
={}&
c\sum_{m'=0}^{\infty}
\int_0^\infty
\left(\prod_{\beta=1}^{m'+1}d\Delta t'_\beta\right)
\delta\!\left(
t-\sum_{\beta=1}^{m'+1}\Delta t'_\beta
\right)
\mathcal Q^{n_i1}
\bigl[\{n_i(\tau)\}_t\mid\{n_{i+1}(\tau)\}_t\bigr]
\Lambda_{m'}^{1}(\Delta\mathbf t'_{m'+1})
\\
&+(1-c)\sum_{m'=0}^{\infty}
\int_0^\infty
\left(\prod_{\beta=1}^{m'+1}d\Delta t'_\beta\right)
\delta\!\left(
t-\sum_{\beta=1}^{m'+1}\Delta t'_\beta
\right)
\mathcal Q^{n_i0}
\bigl[\{n_i(\tau)\}_t\mid\{n_{i+1}(\tau)\}_t\bigr]
\Lambda_{m'}^{0}(\Delta\mathbf t'_{m'+1}).
\label{eq:Paper2_FinalExplicitEqSystem}
\end{aligned}
\end{equation}
\end{widetext}

There are infinitely many unknown functionals in 
Eq.~(\ref{eq:Paper2_FinalExplicitEqSystem}), yet the equations are closed. 
To see this, note that the unknowns on the left-hand side are 
\(\{\Lambda_m^0, \Lambda_m^1\}\) with \(m=0,1,\ldots,\infty\). 
The same set of unknowns appears on the right-hand side, and each unknown 
has a corresponding equation. Thus the number of unknowns matches the 
number of equations, even though both are infinite.

\subsection{Kernel of the path-probability equations}
\label{sec:kernel}
We now consider the form of the kernel
\(\mathcal{Q}^{n_i n_{i+1}}[\{n_i(\tau)\}_t \,|\, \{n_{i+1}(\tau)\}_t]\)
for a given pair of neighboring spin trajectories. Let the trajectory of \(n_i\) be
\(\{n_i(0)=n_i; t_1, t_2, \ldots, t_m, t\}\)
and that of \(n_{i+1}\) be
\(\{n_{i+1}(0)=n_{i+1}; t'_1, t'_2, \ldots, t'_{m'}, t\}\).
An example of such a trajectory pair is shown in Fig.~\ref{fig:Paper2_PathOfTwoSpin}. There is a unique division of the total time interval \([0,t]\) into segments. Within each segment, no flipping events occur for either spin \(i\) or spin \(i+1\). The beginning time \(t_a\) and end time \(t_b\) of a segment are two successive times in the combined sequence
\(0, t_1, t_2, \ldots, t_m\) and \(0, t'_1, t'_2, \ldots, t'_{m'}\).
In general, there are only three types of segment, classified according to the dynamics of \( n_{i+1}\). 
The total kernel is the product of \(Q_{ab}\) over all segments,
\begin{equation}
\mathcal{Q}^{n_i n_{i+1}}[\{n_i(\tau)\}_t | \{n_{i+1}(\tau)\}_t] = \prod_{(t_a, t_b)} Q_{ab} (t_b - t_a),
\label{eq:Q_prod}
\end{equation}
where 
\begin{equation}
Q_{ab} (t_b - t_a) =
\begin{cases}
Q_{ab}^0 = 1, & \text{0-segment}, \\[0pt]
Q_{ab}^{1\text{-}1} = \omega_{n_i}\, e^{-\omega_{n_i}(t_b-t_a)}, & \text{1-1 segment}, \\[0pt]
Q_{ab}^{1\text{-}0} = e^{-\omega_{n_i}(t_b-t_a)}, & \text{1-0 segment}.
\end{cases}
\label{eq:Qab}
\end{equation}
Below, we show the derivation of \(Q_{ab}\) for each type of segment.

{\it (i) \(0\)-segment.} This corresponds to the situation in which \(n_{i+1}=0\) throughout the interval \((t_a,t_b)\), regardless of whether \(n_{i+1}\) flips at \(t_b\).
In this case, spin \(i\) is not allowed to flip, so its state remains unchanged. The corresponding contribution to the kernel is simply
\begin{equation}
    Q_{ab}^{0} = 1.
\end{equation}

{\it (ii) \(1\)-\(1\) segment.}
This corresponds to the case in which \(n_{i+1}=1\) throughout the interval \((t_a,t_b)\), and $n_{i+1}$ does not flip at $t_b$. In this case, spin \(i\) flips at \(t_b\). The probability density for spin \(i\) to undergo its first flip after a time \(t_b-t_a\) is given by Eq.~(\ref{eq:p_flip}):
\begin{equation}
    Q_{ab}^{1\text{-}1}(t_b-t_a)
    = p_{\rm flip}(t_b-t_a)
    = \omega_{n_i}\exp[-\omega_{n_i}(t_b-t_a)].
    \label{eq:Paper2_FirstFlipDensity}
\end{equation}
Here, \(n_i(t_a)=n_i\).

{\it (iii) \(1\)-\(0\) segment.}
This situation occurs when \(n_{i+1}=1\) throughout the interval \((t_a,t_b)\), but \(n_{i+1}\) flips to \(0\) at \(t_b\). Thus, spin \(i\) is facilitated during this interval but does not flip.
The relevant contribution is the probability that spin \(i\) remains unflipped from \(t_a\) to \(t_b\). This is the persistence probability obtained by integrating the flipping-time distribution in Eq.~(\ref{eq:p_flip}):
\begin{equation}
   Q_{ab}^{1\text{-}0}(t_b - t_a) = \int_{t_b - t_a}^\infty p_{\rm flip}(\tau)\,d\tau = \exp[-\omega_{n_i}(t_b - t_a)].
    \label{eq:Paper2_PersistProb}
\end{equation}
Note that, if $n_{i+1}=1$ in the last segment, then this segment should be treated as  \(1\)-\(0\) segment, even though 
$n_{i+1}$ does not flip to $0$ at $t_b=t$.

Having specified the three types, one can write down an explicit expression for any
\(\mathcal{Q}^{n_i n_{i+1}}[\{n_i(\tau)\}_t | \{n_{i+1}(\tau)\}_t]\)
using Eq.~(\ref{eq:Q_prod}). The expression for the example in Fig.~\ref{fig:Paper2_PathOfTwoSpin} is given in the figure caption.

Thus far, the path-probability equations Eq.~(\ref{eq:Paper2_FinalExplicitEqSystem})  have been formally established. 
Solving this infinite set of equations is extremely challenging. 
If this system can be solved, one can use 
Eq.~\eqref{eq:Paper2_P11SchematicCal} to obtain the return probability 
and then Eq.~\eqref{eq:Paper2_CorrFunc} to obtain the time correlation 
function. In what follows, we discuss our efforts to solve these equations; we focus on equilibrium dynamics and omit the superscript ``eq'' from the 
notation.

\section{Laplace-domain formulation}

\subsection{Path-probability equations}
The derivation of the path-probability equations in the Laplace domain is explicit, albeit lengthy; see Appendices~\ref{app:first} and~\ref{app:second} for details.
The path-probability equations Eq.~(\ref{eq:Paper2_FinalExplicitEqSystem}) become:
\begin{widetext}
    \begin{equation}
        \begin{split}
            \tilde{\Lambda}_{m}^{n_i}({\bf z}_{m+1})=(-1)^{m}\omega_{n_i}^{\lceil m/2\rceil}\omega_{\bar{n}_i}^{\lfloor m/2\rfloor}\sum_{m'=0}^{\infty}\sum_{\{x_{\alpha}\}}
            \delta \left(m - \sum_{\alpha=1}^{m'+1}x_\alpha \right)
            \prod_{\alpha=1}^{m'+1}\mathcal{S}_{\alpha,x_{\alpha}}^{\{\hat{z}_{\alpha}^{(\gamma_{\alpha})}\}}\Bigg[c\tilde{\Lambda}_{m'}^{1}({\bf z}'_{m'+1})+(1-c)\tilde{\Lambda}_{m'}^{0}({\bf z}'_{m'+1})\Bigg].
        \end{split}
            \label{eq:Paper2_FinalLaplaceExpression}
    \end{equation}
\end{widetext}

Here, \(\tilde{\Lambda}_{m}^{n_i}(\mathbf{z}_{m+1})\) is the Laplace transform of \(\Lambda_{m}^{n_i}(\Delta \mathbf{t}_{m+1})\), and \(\mathbf{z}_{m+1}=\{z_1,z_2,\ldots,z_{m+1}\}\) are the Laplace variables conjugate to \(\Delta \mathbf{t}_{m+1}\). Below, we use tilded notation to denote the Laplace transform of a function. In \(\tilde{\Lambda}_{m}^{n_i}(\mathbf{z}_{m+1})\), \(m\) denotes the total number of flips of spin \(i\), and \(n_i\) is the initial condition, \(n_i(t=0)=n_i\). Similarly, in \(\tilde{\Lambda}_{m'}^{0}(\mathbf{z}'_{m'+1})\), \(m'\) is the total number of flips of spin \(i+1\), and \(0\) is the initial condition, \(n_{i+1}(t=0)=0\).
In the prefactor
\(
\omega_{n_i}^{\lceil m/2\rceil}\omega_{\bar n_i}^{\lfloor m/2\rfloor},
\)
\(\lceil \cdot \rceil\) and \(\lfloor \cdot \rfloor\) denote the ceiling and floor functions, respectively.

In the summation \(\sum_{\{x_\alpha\}}'\), 
\(\{x_\alpha\}\) with \(1\leq \alpha \leq m'+1\) indicates that spin \(i\) flips \(x_\alpha\) times during the time interval \((t'_{\alpha-1},t'_{\alpha})\), where \(t'_0=0\). Note that we use \(t_\alpha\) and \(t'_\alpha\) to denote flipping times for spin \(i\) and \(i+1\), respectively. 
The delta function denotes that we sum over all \(\{x_{\alpha}\}\) under the constraint
\begin{equation}
\sum_{\alpha=1}^{m'+1} x_\alpha=m.
\label{eq:sum_x_alpha}
\end{equation}

In Eq.~(\ref{eq:Paper2_FinalLaplaceExpression}), \(\mathcal{S}_{\alpha,x_{\alpha}}^{\{\hat{z}_{\alpha}^{(\gamma_{\alpha})}\}}\) is a kernel operator whose expression, although complicated, is explicit.
This operator acts on \(\tilde{\Lambda}_{m'}^{0}(\mathbf{z}'_{m'+1})\) or \(\tilde{\Lambda}_{m'}^{1}(\mathbf{z}'_{m'+1})\), over the interval \(\Delta t'_\alpha\) of spin \(i\), conjugate to the argument \(z_\alpha\).
During \(\Delta t'_\alpha\), \(x_\alpha\) flips of spin \(i\) occur, corresponding to the variables
\(
\hat{z}_\alpha^{(1)}, \hat{z}_\alpha^{(2)}, \dots, \hat{z}_\alpha^{(\gamma_\alpha)}, \ldots, \hat{z}_\alpha^{(x_\alpha+1)},
\)
with \(1 \leq \gamma_\alpha \leq x_\alpha+1\).
A detailed derivation of \(\mathcal{S}_{\alpha,x_{\alpha}}^{\{\hat{z}_{\alpha}^{(\gamma_\alpha)}\}}\) is provided in Appendix~\ref{app:second}; here we outline the result. 

\begin{figure}
    \centering
    \includegraphics[width=\linewidth]{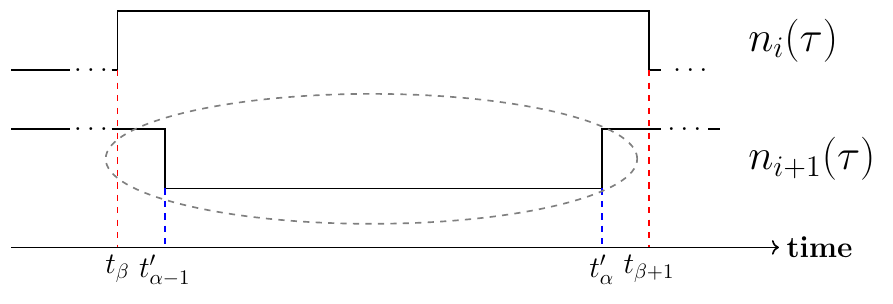}
    \caption{Schematic illustration of the case when spin $i$ is not allowed to flip during a $0$-segment of spin $i+1$.
    The $0$-segment is highlighted by the dashed ellipse. }
    \label{fig:Paper2_0Segment}
\end{figure}

\textit{(i) \(n_{i+1}=0\) throughout \(\Delta t'_{\alpha}\) and spin $i$ is not allowed to flip.}
Owing to the dynamical constraint, spin \(i\) cannot flip during the \(0\)-segments of spin \(i+1\), and thus $x_\alpha$ has to be zero (see Fig.~\ref{fig:Paper2_0Segment}).
We thus adopt the convention that, 
whenever \(x_{\alpha}>0\) and the interval \(\Delta t'_{\alpha}\) over which \(\mathcal{S}_{\alpha,x_{\alpha}}^{\{\hat{z}_{\alpha}^{(\gamma_\alpha)}\}}\) acts is a \(0\)-segment, the result is taken to be \(0\):
\begin{equation}
\mathcal{S}_{\alpha,x_{\alpha}}^{\{\hat{z}_{\alpha}^{(\gamma_\alpha)}\}}
\tilde{\Lambda}_{m'}^{n_{i+1}}(\mathbf{z}'_{m'+1})=0,
\quad
\text{if } n_{i+1}(t)=0 \text{ on } \Delta t'_{\alpha}.
\end{equation}
For example, $ \mathcal{S}_{2,1}^{\{\hat{z}_2^{(1)},\hat{z}_2^{(2)}\}} \tilde{\Lambda}_1^1(z'_1,z'_2) =
\mathcal{S}_{1,1}^{\{\hat{z}_1^{(1)},\hat{z}_1^{(2)}\}}
\tilde{\Lambda}_1^{0}(z'_1,z'_2)=0$.

\begin{figure}
    \centering
    \includegraphics[width=\linewidth]{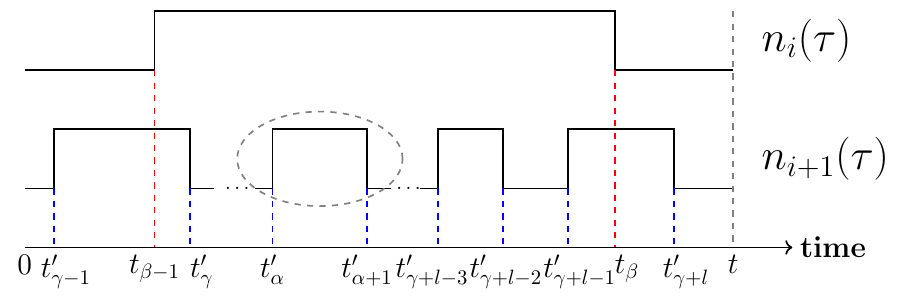}
    \caption{Schematic illustration of the case when spin \(i+1\) flips multiple times between two consecutive flips of spin \(i\).
    The dashed ellipse indicates an example of the situation (ii) in the discussion of the kernel operator: there is no flip of spin \(i\) during \(\Delta t'_{\alpha}\).
    }
    \label{fig:Paper2_Trajectory01NoFlipPlateau}
\end{figure}

\textit{(ii) \(n_{i+1}=1\) throughout \(\Delta t'_{\alpha}\); spin $i$ is allowed to flip but does not.}
In this case (see Fig.~\ref{fig:Paper2_Trajectory01NoFlipPlateau}), \(x_\alpha=0\). Suppose the interval \(\Delta t'_{\alpha}\) of spin \(i+1\) lies within an interval \(\Delta t_\beta\) of spin \(i\) between two flips, during which spin \(i\) is in state \(n_i\). The Laplace variable in \(\tilde{\Lambda}_{m}^{n_i}(\mathbf{z}_{m+1})\) conjugate to \(\Delta t_\beta\) is \(z_\beta\). Then the operator is
\begin{equation}
\mathcal{S}_{\alpha,0}^{\hat{z}_\alpha}
\tilde{\Lambda}_{m'}^{n_{i+1}}(\mathbf{z}'_{m'+1})
=
\tilde{\Lambda}_{m'}^{n_{i+1}}(\ldots, \hat{z}_\alpha, \ldots),
\end{equation}
where
\begin{equation}
\hat{z}_\alpha=
\begin{cases}
z_\beta, & \text{for a \(0\)-segment},\\[2pt]
z_\beta+\omega_{n_i}, & \text{for a \(1\)-\(0\) segment}.
\end{cases}
\label{eq:hatz_main}
\end{equation}
Thus, the operator copies the argument $z_\beta$ or shifts it by $\omega_{n_i}$.

\begin{figure}
    \centering
    \includegraphics[width=\linewidth]{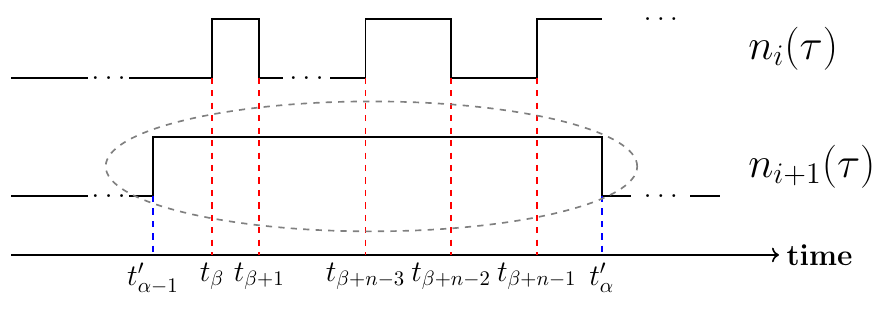}
    \caption{Schematic illustration of a $1$-plateau containing multiple flips of spin $i$. The region enclosed by the dashed ellipse is a $1$-plateau containing multiple flip events of spin $i$.}
    \label{fig:Paper2_TrajectoryMultiFlip}
\end{figure}

\textit{(iii) \(n_{i+1}=1\) throughout \(\Delta t'_{\alpha}\); spin $i$ is allowed to flip and flips one or more times.}
Suppose that during the interval \(\Delta t'_{\alpha}\) of \(n_{i+1}=1\), one or more flips of spin \(i\) occur (see Fig.~\ref{fig:Paper2_TrajectoryMultiFlip}). 
We call such an interval \(\Delta t'_{\alpha}\)  a {\it 1-plateau}.
The first flip of $i$ occurs at \(t_\beta\) within \(\Delta t'_{\alpha}\).
The interval \(\Delta t_\beta\), during which spin \(i\) is in state \(n_i\), is conjugate to \(z_\beta\). Then
\begin{equation}
\begin{split}
&\mathcal{S}_{\alpha,x_{\alpha}}^{\{\hat{z}_{\alpha}^{(\gamma_{\alpha})}\}}
\tilde{\Lambda}_{m'}^{n_{i+1}}(\ldots, z'_\alpha, \ldots) \\
&=\left[\hat{z}_{\alpha}^{(1)}, \hat{z}_{\alpha}^{(2)}, \ldots, \hat{z}_{\alpha}^{(x_\alpha+1)} \right] \tilde{\Lambda}_{m'}^{n_{i+1}}(\ldots, z'_\alpha, \ldots)\\
&=\sum_{j=1}^{x_\alpha+1}
\frac{\tilde{\Lambda}_{m'}^{n_{i+1}}(\ldots, \hat{z}_\alpha^{(j)}, \ldots)}
{\prod_{k\ne j}^{x_\alpha+1}\left(\hat{z}_{\alpha}^{(j)}-\hat{z}_{\alpha}^{(k)} \right)}.
\end{split}
\end{equation}
where the operator \(\left[\hat{z}_{\alpha}^{(1)}, \hat{z}_{\alpha}^{(2)}, \ldots, \hat{z}_{\alpha}^{(x_\alpha+1)} \right]\) is the divided difference of order \(x_\alpha\) acting on the \(\alpha\)-th argument of \(\tilde{\Lambda}_{m'}^{n_{i+1}}\), and
\begin{equation}
    \hat{z}_{\alpha}^{(\gamma_\alpha)}=\begin{cases}
        z_{\beta+\gamma_\alpha-1}+\omega_{n_i}, & \gamma_\alpha\text{ is odd},\\[2pt]
        z_{\beta+\gamma_\alpha-1}+\omega_{\bar{n}_i}, & \gamma_\alpha \text{ is even}.
        \label{eq:Paper2_hatzMultiFlip}
    \end{cases}
\end{equation}

As in Eq.~\eqref{eq:Paper2_FinalExplicitEqSystem}, the Laplace-domain equation Eq.~\eqref{eq:Paper2_FinalLaplaceExpression} is complete and closed, but it consists of an infinite number  of equations indexed by \(m=0,1,2,\dots\). It converts the multiple integral in Eq.~\eqref{eq:Paper2_FinalExplicitEqSystem} into an algebraic form, but solving it remains extremely challenging. Below, we discuss an approximation scheme for solving these equations.

\section{Factorization Approximation Scheme}

\subsection{Systematic factorization approximation}

Equation~\eqref{eq:Paper2_FinalLaplaceExpression} represents an infinite family of equations indexed by \(m = 0,1,2,\ldots\) (i.e., \(0 \leq m < \infty\)). The most natural approximation is to factorize the path-probability density into a product of lower-order path-probability densities. As defined in Eq.~\eqref{eq:Paper2_PathProbDensityExplicit}, \(\Lambda_m^{n_i}(\Delta \mathbf{t}_{m+1})\) is simply the joint probability of \(m+1\) flip intervals. To illustrate the factorization approximation, we consider a general joint probability distribution \(P(\Delta t_1, \Delta t_2, \ldots, \Delta t_{m+1})\).

In the first-order approximation, we factorize it into a product of single-interval probability densities:
\begin{equation}
P(\Delta t_1, \Delta t_2, \ldots, \Delta t_{m+1})
\approx
P(\Delta t_1)\,P(\Delta t_2)\cdots P(\Delta t_{m+1}).
\label{eq:first-order}
\end{equation}
In the second-order approximation,
\begin{equation}
P(\Delta t_1, \Delta t_2, \ldots, \Delta t_{m+1})
\approx
P(\Delta t_1, \Delta t_2)\,P(\Delta t_3, \Delta t_4)\cdots.
\end{equation}
This scheme generalizes to arbitrary \(k\)-th order. In general, the \(k\)-th order approximation assumes that decorrelation occurs after every \(k\) consecutive flips of spin \(i\). The factorization expressions above are schematic. Unlike the preceding intervals, each of which terminates at a spin-flip event, the final interval ends at the observation time without requiring an additional flip. Its contribution is therefore described by a persistence probability rather than a next-flip-time density. The explicit factorization formulas account for this distinction.

Our factorization scheme is closely analogous to the well-known Kirkwood superposition approximation, which is used to factorize spatial correlations of particle distributions in fluids~\cite{kirkwood1935statistical}. 
For instance, the Kirkwood approximation factorizes the three-body correlation function into two-body correlation functions as
$g_3(r_1,r_2,r_3)
\sim
g_2(r_1,r_2)\,g_2(r_2,r_3)\,g_2(r_1,r_3)$.
In our case, however, we do not need to include terms such as \(P(\Delta t_1, \Delta t_3)\), because \(\{\Delta t_1,\Delta t_2,\ldots,\Delta t_{m+1}\}\) is a one-dimensional ordered sequence.

When this factorization scheme is applied to \(\Lambda_m^{n_i}(\Delta \mathbf{t}_{m+1})\), extra care is required because of the kinetic constraint rules. We discuss this in detail below.


\subsection{Single-interval factorization}

\begin{figure}
    \centering
    \includegraphics[width=\linewidth]{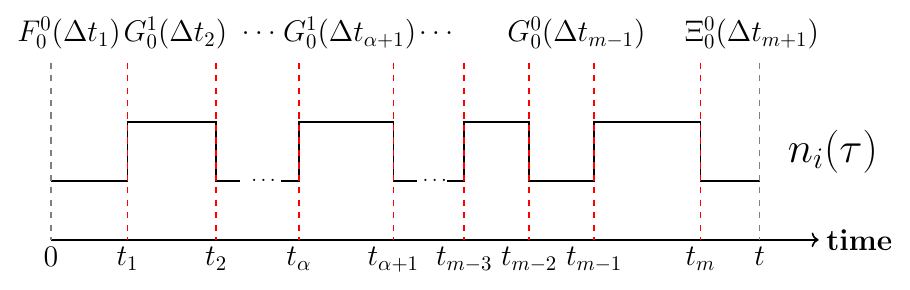}
    \caption{Single-interval factorization. An example of the path decomposition for spin $i$, where the functions labeled above the trajectory correspond to the 1-IF given by Eq.~\eqref{eq:Paper2_1stDecompOfLambdaXi}.
    }
    \label{fig:placeholder}
\end{figure}

Applying the first-order factorization Eq.~\eqref{eq:first-order}, we approximate \(\tilde{\Lambda}_m^{n_i}(\mathbf{z}_{m+1})\) by a product of single-interval probabilities, which, in the Laplace domain, is written as (see Fig.~\ref{fig:placeholder}):
\begin{equation}
  \tilde{\Lambda}_m^{n_i}(\mathbf{z}_{m+1})
  =
  \tilde{F}_0^{n_i}(z_1)
  \prod_{\alpha=2}^{m}\Big[\tilde{G}_0^{\delta_{\alpha}^{n_i}}(z_\alpha)\Big]
  \tilde{\Xi}_0^{\delta_{m+1}^{n_i}}(z_{m+1}).
  \label{eq:Paper2_1stDecompOfLambdaXi}
\end{equation}
Here, \(\tilde{F}_0^{n_i}(z_1)\) is the Laplace transform of \(F_0^{n_i}(\Delta t_1)\), associated with the first segment. It denotes the \textit{next-flip-time density}, i.e., the probability density that the next flip (here the first flip) occurs at time \(\Delta t_1\), with no additional constraints.
The relationship between \(F_0^{n_i}\) and \(\Lambda_{0}^{n_i}\) can be derived as follows. The path-probability density \(\Lambda_0^{n_i}(\Delta t_1)\) describes trajectories in which spin \(i\) undergoes zero flips during \(\Delta t_1 = t - 0\), and the first flip can occur at any time after \(t\). Thus, \(\Lambda_{0}^{n_i}(\Delta t)\) can be obtained by integrating the next-flip-time density \(F_0^{n_i}(\Delta \tau)\) over all possible first-flip times \(\Delta \tau > \Delta t\), yielding
\begin{equation}
    \Lambda_{0}^{n_i}(\Delta t)
    =
    \int_{\Delta t}^{\infty} d \Delta \tau \, F_0^{n_i}(\Delta \tau).
    \label{eq:Paper2_Lambda0F0}
\end{equation}

For the intermediate segments (from the second to the penultimate), the probabilities are described by \(\tilde{G}_0^{\delta_{\alpha}^{n_i}}(\Delta t_\alpha)\), where \(\delta_\alpha^{n_i}=n_i\) for odd \(\alpha\) and \(\delta_\alpha^{n_i}=\bar{n}_i\) for even \(\alpha\). This is the conditional next-flip-time density, with the constraint that spin \(i+1\) must be in state 1 at \(t_{\alpha-1}\), because the previous flip of spin \(i\) occurs exactly at \(t_{\alpha-1}\). Correspondingly, we introduce a conditional path-probability:
\begin{equation}
    \Xi_{m}^{n_i}(\Delta \mathbf{t}_{m+1})
    =
    \mathcal{P}_i[\{n_i(0);\mathbf{t}_m,t\}\,|\,n_{i+1}(0)=1].
    \label{eq:Paper2_AuxliaryPPD}
\end{equation}
When \(m=0\), we have
\begin{equation}
    \Xi_{0}^{n_i}(\Delta t)
    =
    \int_{\Delta t}^{\infty} d\Delta \tau \, G_0^{n_i}(\Delta \tau).
    \label{eq:Paper2_RelationOfPath&NextFlipDensityG0}
\end{equation}
It is easy to see that the probability associated to the last segment is 
 $\Xi_{0}^{\delta^{n_i}_{m+1}}(\Delta t_{m+1})$.

Laplace transformation of Eqs.~(\ref{eq:Paper2_Lambda0F0}) and (\ref{eq:Paper2_RelationOfPath&NextFlipDensityG0}) gives,
\begin{equation}
    \begin{split}
        &\tilde{F}_0^{n_i}(z)=1-z\tilde{\Lambda}_0^{n_i}(z),\\
        &\tilde{G}_0^{n_i}(z)=1-z\tilde{\Xi}_0^{n_i}(z).
    \end{split}
    \label{eq:Paper2_RelationOfPath&NextFlipDensity_nEqual0}
\end{equation}

\subsection{Closed equations under the single-interval factorization approximation}

Under the 1-IF in Eq.~\eqref{eq:Paper2_1stDecompOfLambdaXi}, for $m'\geq 1$, $\Lambda_{m'}^{n_{i+1}}(\Delta {\bf t}'_{m'+1})$ can be recursively factorized into products of $F_0^{n_{i+1}}(\Delta t'_1)$, $G_0^{\bar{n}_{i+1}}(\Delta t_2')$, $G_0^{n_{i+1}}(\Delta t_{3})$, $\ldots$ and $\Xi_0^{n_{i+1}}(\Delta t'_{m'+1})$(or $\Xi_0^{\bar{n}_{i+1}}(\Delta t'_{m+1})$).
Thus, only the \(m=0\) quantities remain as independent unknowns. This greatly simplifies the 
path-probability equations: as shown in this subsection, we obtain four equations 
for four unknown probabilities, which can be solved numerically.

First, setting $m=0$ in Eq.~\eqref{eq:Paper2_FinalLaplaceExpression}, we obtain
\begin{widetext}
    \begin{equation}
        \begin{split}
            \tilde{\Lambda}_0^{n_i}(z_1)&=\sum_{m'=0}^{\infty}\prod_{\alpha=1}^{m'+1}\mathcal{S}_{\alpha,0}^{\hat{z}_{\alpha}}\Bigg[c\tilde{\Lambda}_{m'}^{1}({\bf z}'_{m'+1})+(1-c)\tilde{\Lambda}_{m'}^{0}({\bf z}'_{m'+1})\Bigg]\\
            &=\sum_{m'=0}^{\infty}\Bigg[c\tilde{\Lambda}_{m'}^{1}(\hat{\bf z}_{m'+1})+(1-c)\tilde{\Lambda}_{m'}^0(\hat{\bf z}_{m'+1})\Bigg],
        \end{split}
    \label{eq:Paper2_FinalLaplace_m=0}
\end{equation}
\end{widetext}
Because of the constraint Eq.~(\ref{eq:sum_x_alpha}) and $m=0$,  we require that $x_\alpha =0$ for any $\alpha$.
In Eq.~(\ref{eq:Paper2_FinalLaplace_m=0}), $\hat{\bf z}_{m'+1}=\{\hat{z}_1,\hat{z}_2,...,\hat{z}_{m'+1}\}$, with each $\hat{z}_{\alpha}$ given by Eq.~\eqref{eq:hatz_main}. Specifically, for $\tilde{\Lambda}_{m'}^1(\hat{\bf z}_{m'+1})$, we have $\hat{\bf z}_{m'+1}=\{z_1+\omega_{n_i},z_1, z_1+\omega_{n_i}, \ldots\}$,  whereas for $\tilde{\Lambda}_{m'}^0(\hat{\bf z}_{m'+1})$, we have $\hat{\bf z}_{m'+1}=\{z_1,z_1+\omega_{n_i}, z_1, \ldots\}$. The two terms $z_1$ and $z_1+\omega_{n_i}$ continue to alternate throughout the sequence.

Equation~(\ref{eq:Paper2_FinalLaplace_m=0}) relates $\tilde{\Lambda}_0^{n_i}(z_1)$ of spin $i$ to $\tilde{\Lambda}_{m'}^0(\hat{\bf z}_{m'+1})$ and $\tilde{\Lambda}_{m'}^1(\hat{\bf z}_{m'+1})$ of spin $i+1$. Similarly, we can relate 
$\tilde{\Xi}_0^{n_i}$ to probabilities of spin $i+1$.
Since $\tilde{\Xi}_{0}^{n_i}(z_1)$ is the path-probability density for trajectories in which spin $i+1$ is initially in state $1$, it is related only to  $\tilde{\Lambda}_{m'}^1(\hat{\bf z}_{m'+1})$, as:
\begin{equation}
        \tilde{\Xi}_0^{n_i}(z_1)=\sum_{m'=0}^{\infty}\tilde{\Lambda}_{m'}^{1}(\hat{\bf z}_{m'+1}).
\label{eq:Paper2_FinalLaplace_m=0_Xi}
\end{equation}

To close Eqs.~\eqref{eq:Paper2_FinalLaplace_m=0} and~\eqref{eq:Paper2_FinalLaplace_m=0_Xi}, we factorize \(\Lambda_{m'}^{n_i}\) on the right-hand side into probabilities with \(m=0\), using Eq.~\eqref{eq:Paper2_1stDecompOfLambdaXi}.
Furthermore, we eliminate two unknown probabilities via Eq.~\eqref{eq:Paper2_RelationOfPath&NextFlipDensity_nEqual0}. 
Finally, Eqs.~\eqref{eq:Paper2_FinalLaplace_m=0} and~\eqref{eq:Paper2_FinalLaplace_m=0_Xi} become closed as follows:
\begin{widetext}
\begin{subequations}
\begin{align}
    \tilde{\Lambda}_0^{n_i}(z)&=\frac{\begin{aligned}\tilde{\Xi}_0^0(z)+\tilde{\Xi}_{0}^1(z+\omega_{n_i})-[z&+(1-c)\omega_{n_i}]\tilde{\Xi}_{0}^0(z)\tilde{\Xi}_{0}^1(z+\omega_{n_i})\\
    &+\omega_{n_i}[(1-c)\tilde{\Lambda}_0^0(z)\tilde{\Xi}_{0}^1(z+\omega_{n_i})-c\tilde{\Xi}_{0}^0(z)\tilde{\Lambda}_0^1(z+\omega_{n_i})]\end{aligned}}{z\tilde{\Xi}_{0}^0(z)+(z+\omega_{n_i})\tilde{\Xi}_{0}^1(z+\omega_{n_i})-z(z+\omega_{n_i})\tilde{\Xi}_{0}^0(z)\tilde{\Xi}_{0}^1(z+\omega_{n_i})},
\label{eq:Paper2_1stCloseRecusion1}\\
\nonumber
\\
\tilde{\Xi}_{0}^{n_i}(z)&=\frac{\tilde{\Xi}_{0}^0(z)+\tilde{\Xi}_{0}^1(z+\omega_{n_i})-z\tilde{\Xi}_{0}^0(z)\tilde{\Xi}_{0}^1(z+\omega_{n_i})-\omega_{n_i}\tilde{\Xi}_{0}^0(z)\tilde{\Lambda}_{0}^1(z+\omega_{n_i})}{z\tilde{\Xi}_{0}^0(z)+(z+\omega_{n_i})\tilde{\Xi}_{0}^1(z+\omega_{n_i})-z(z+\omega_{n_i})\tilde{\Xi}_{0}^0(z)\tilde{\Xi}_{0}^1(z+\omega_{n_i})},
\label{eq:Paper2_1stCloseRecusion2}
\end{align}
\end{subequations}
\end{widetext}
In total, there are four equations,  and four unknowns $\tilde{\Lambda}_0^{n_i}$ and $\tilde{\Xi}_0^{n_i}$, 
with $n_i = 0$ or 1.

In Appendix~\ref{sec:two-interval}, we extend the factorization scheme to the two-interval level. Although explicit closed path-probability equations are obtained, solving them remains infeasible.

\begin{figure*}[t] %
    \centering
    \includegraphics[width=\textwidth]{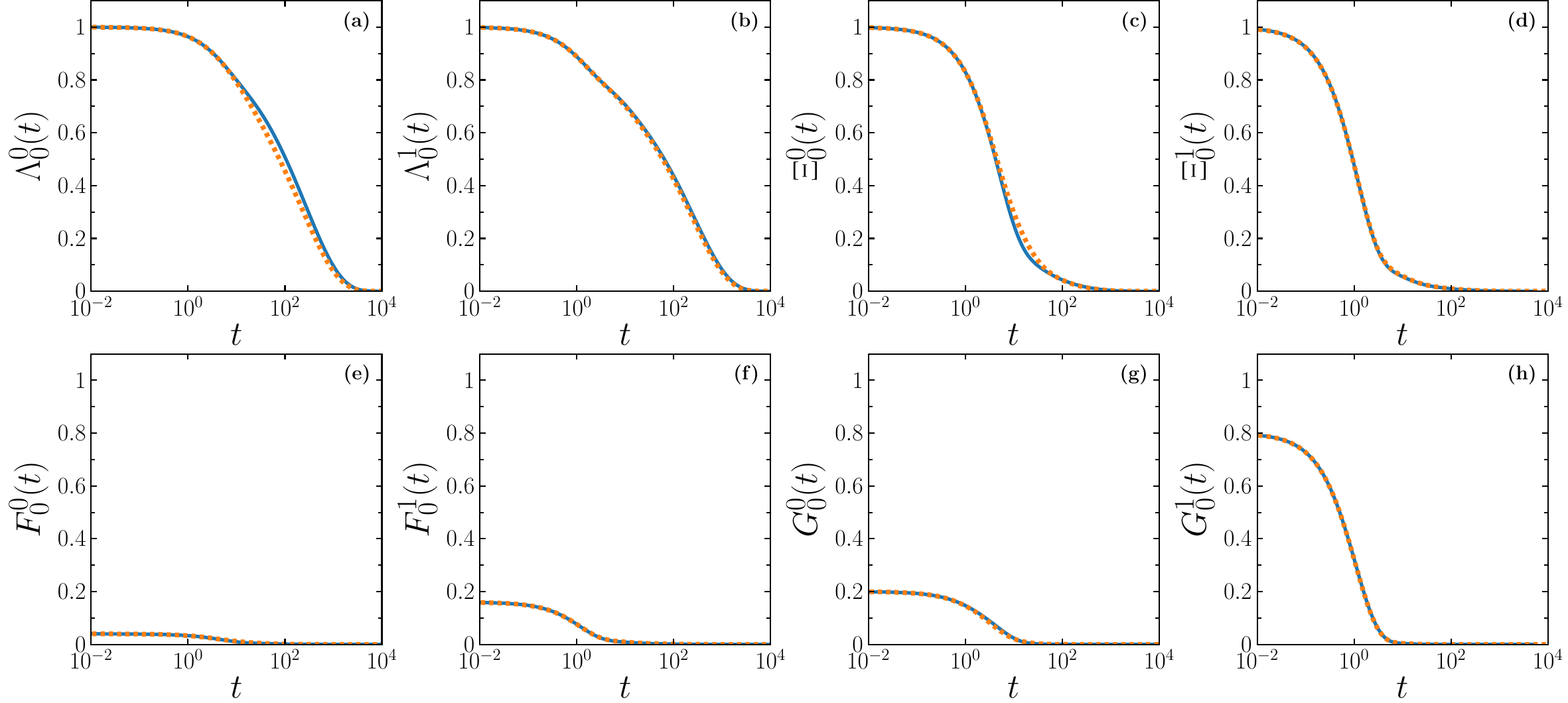} 
    \caption{The figure shows four persistence functions $\Lambda_0^0(t)$, $\Lambda_0^1(t)$, $\Xi_0^0(t)$ and $\Xi_0^1(t)$ in (a-d), together with their corresponding next-flip-time densities in (e-h), for $c=0.2$. Solid lines represent 1-IF results; dashed lines are calculated using the 2-DB approximation based on the Zwanzig–Mori projection-operator
formalism (see Eq.~\eqref{eq:Paper2_PersistToCorrs}).
}
    \label{fig:Paper2_PersistenceCompare}
\end{figure*}

\begin{figure}
    \centering
    \includegraphics[width=\linewidth]{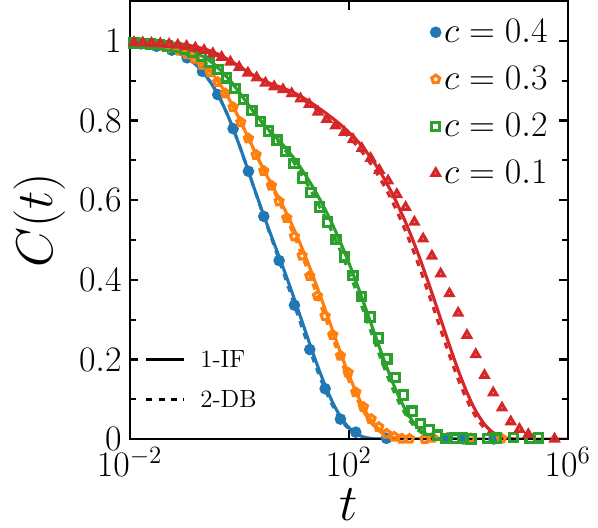}
    \caption{Comparison of the correlation function $C(t)$ predicted by the 1-IF theory, the 2-DB theory~\cite{van2005glassy},  and Monte Carlo simulations. }
    \label{fig:Paper2_CtCompareSIFVanZon}
\end{figure}

\begin{figure}
    \centering
    \includegraphics[width=\linewidth]{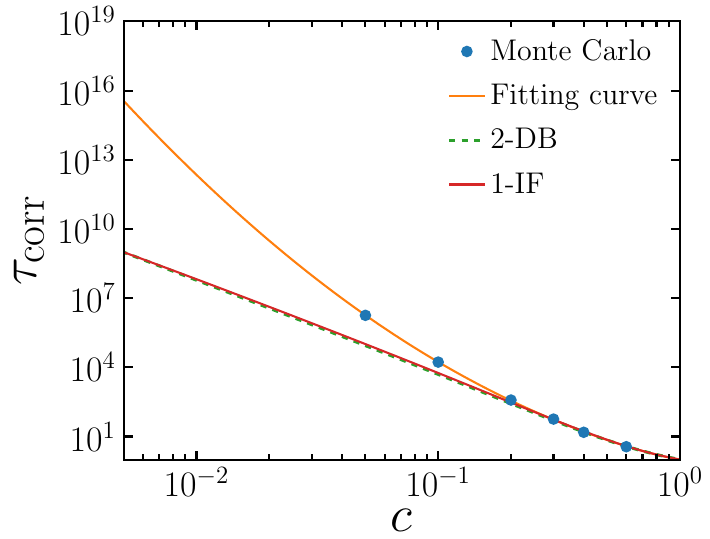}
    \caption{The relaxation time $\tau_{\rm corr}$ as a function of the mean occupancy $c$. The Monte Carlo simulation result is fitted by an empirical function $\frac{1}{c}\exp{\big[\frac{(\ln{c})^2}{2\ln{2}}+A\ln{\frac{1}{c}}+B\ln{(1+\ln{\frac{1}{c}})}+D(1-c)\big]}$, where $A\approx3.98$, $B\approx-9.04$ and $D\approx5.83$. 
    This empirical form matches the exact scaling Eq.~(\ref{eq:tau}) in the limit $c \to 0$.
    }
    \label{fig:TauCompareAll}
\end{figure}

\subsection{Numerical solution under the single-interval factorization approximation and comparison with simulations and other theories}

Equations~\eqref{eq:Paper2_1stCloseRecusion1} and~\eqref{eq:Paper2_1stCloseRecusion2} do not admit an analytical treatment. We therefore solve them numerically using an iterative scheme, carrying out the iteration up to a truncation level of \(128\) iterations, at which point the relaxation time is found to have converged. Figure~\ref{fig:Paper2_PersistenceCompare}(a--d) shows the solutions of $\Lambda_0^{n_i}(t)$ and $\Xi_0^{n_i}(t)$ for $c=0.2$, with $n_i=0$ or $1$. 
Figure~\ref{fig:Paper2_PersistenceCompare}(e--h) shows the corresponding next-flip-time densities (see Eq.~\ref{eq:Paper2_RelationOfPath&NextFlipDensity_nEqual0}).
We then compute \(P[n_i(t)=1 \mid n_i(0)=1]\) in Eq.~\eqref{eq:Paper2_LTOfCorrToTraj} by inverse Laplace transform, using the 1-IF for \(\tilde{\Lambda}_{2m}^1\). Finally, the correlation function is obtained via Eq.~\eqref{eq:Paper2_CorrFunc}.

To test the theoretical results, we perform continuous-time Monte Carlo simulations to obtain the correlation functions for \(c=0.4, 0.3, 0.2\), and \(0.1\)~\cite{newman1999monte}. The results from the 1-IF theory and the Monte Carlo simulations are compared in Fig.~\ref{fig:Paper2_CtCompareSIFVanZon}. The agreement holds down to \(c=0.2\), while for \(c=0.1\), the discrepancy between theory and simulation remains small.

\begin{figure}
    \centering
    \includegraphics[width=\linewidth]{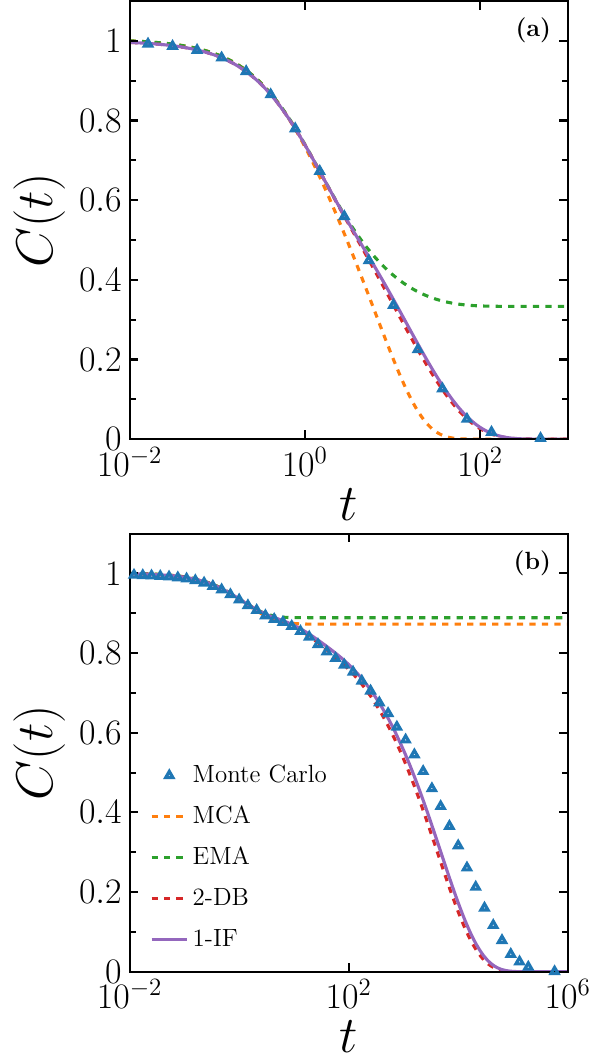}
    \caption{Results of correlation functions in {\bf (a)} $c=0.4$ and {\bf (b)} $c=0.1$.
    }
    \label{fig:Paper2_AllCtCompare}
\end{figure}

It is interesting to compare our theory with previous theories based on the Zwanzig–Mori projection-operator formalism (see our review~\cite{Zha2026Dynamical} for applications of this formalism to the East model). Although exact solutions remain out of reach, various approximation schemes have been developed within that formalism. The best tractable theoretical results so far are obtained using the domain basis proposed by van Zon and Schofield, truncated at the second level~\cite{van2005glassy} (the 2-DB theory). Remarkably, the correlation functions obtained using our 1-IF theory  appear to be nearly identical to the 2-DB results for all studied values of \(c\) (see Fig.~\ref{fig:Paper2_CtCompareSIFVanZon}).
To confirm this, we plot the relaxation time $\tau_{\rm corr}(c)$ obtained from the 1-IF theory, the 2-DB theory, and Monte Carlo simulations in Fig.~\ref{fig:TauCompareAll}. This comparison shows that the results from 1-IF and  2-DB theories are remarkably similar. To the best of our knowledge, these two theories currently show the best agreement with simulations for correlation functions at low $c$ (see Figs.~\ref{fig:Paper2_CtCompareSIFVanZon} and~\ref{fig:Paper2_AllCtCompare}, as well as the discussion in Ref.~\cite{Zha2026Dynamical}). However, as shown in Fig.~\ref{fig:TauCompareAll}, both theories fail to capture the correct asymptotic scaling of $\tau_{\rm corr}(c)$ in Eq.~\eqref{eq:tau}. 

To examine the consistency between the two approaches at the level of the persistence functions, we compare the direct solutions of the 1-IF closure equations with those reconstructed from the correlation functions obtained using the 2-DB approximation (see Appendix~\ref{eq:comp_SIF_2domain}). Specifically, we use the relations derived under the 1-IF assumption in Eq.~\eqref{eq:Paper2_PersistToCorrs} to reconstruct the four persistence functions from the 2-DB correlation functions, and compare them with the direct 1-IF results (see Fig.~\ref{fig:Paper2_PersistenceCompare}). For $\Lambda_0^1(t)$ (Fig.~\ref{fig:Paper2_PersistenceCompare}b) and $\Xi_0^1(t)$ (Fig.~\ref{fig:Paper2_PersistenceCompare}d), the two sets of results are almost indistinguishable. For $\Lambda_0^0(t)$ (Fig.~\ref{fig:Paper2_PersistenceCompare}a) and $\Xi_0^0(t)$ (Fig.~\ref{fig:Paper2_PersistenceCompare}c), the results also agree closely at short and long times, while small deviations emerge at intermediate times. 
These differences indicate that the consistency between the two computational routes is not uniform across all four persistence functions. 
The origin of these deviations remains to be clarified in future studies.

In addition, we compare our results with other well-known approximation schemes within the Zwanzig–Mori formalism, including the mode-coupling approximation (MCA) of Kawasaki~\cite{kawasaki1995irreducible} and the effective-medium approximation (EMA) of Eisinger and Jäckle~\cite{eisinger1993analytical}. It is clear that the MCA and EMA give incorrect predictions of a nonvanishing plateau in the correlation function at small \(c=0.1\) (see Fig.~\ref{fig:Paper2_AllCtCompare}b). This plateau is absent in the present 1-IF theory, indicating that 1-IF does not predict an unphysical ergodicity-breaking transition at a finite \(c\), unlike MCA and EMA.
Further details of the previous theories based on the Zwanzig–Mori formalism can be found in Ref.~\cite{Zha2026Dynamical}.

\subsection{Short spin chains}

\begin{figure}
    \centering
    \includegraphics[width=\linewidth]{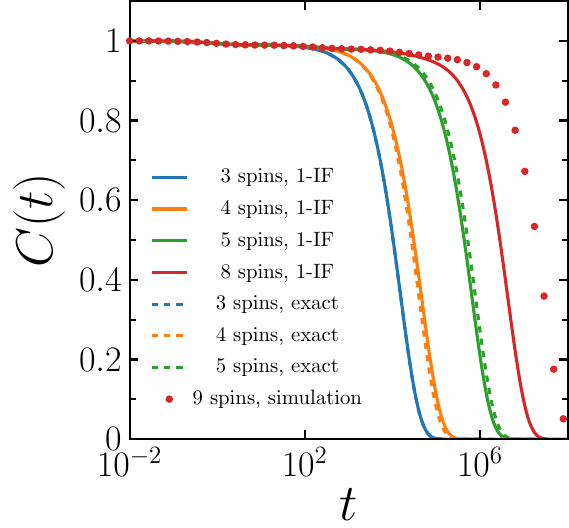}
    \caption{Comparison of the correlation function $C(t)$ obtained from the 1-IF approximation, exact solutions and Monte Carlo simulations for chains of different lengths at $c=0.01$ under free boundary conditions. The 1-IF predictions agree closely with the exact solutions for chains of $3$, $4$ and $5$ spins. For the $9$-spin chain, where obtaining the exact solution becomes computationally demanding, Monte Carlo simulations are used instead, revealing noticeable deviations from the 1-IF prediction, particularly at long times.}
    \label{fig:ShortChain}
\end{figure}

The 1-IF approximation can be reformulated to calculate correlation functions in finite chains. For sufficiently short chains, the 1-IF approximation yields reasonably good results. Due to the finite chain length, spatial translational invariance no longer holds among different spins. Consequently, Eqs.~\eqref{eq:Paper2_1stCloseRecusion1} and ~\eqref{eq:Paper2_1stCloseRecusion2} become a set of recursive relations within the 1-IF approximation for finite chains: the quantities on the left-hand side are the persistence functions of an $N+1$-spin chain, while those on the right-hand sides are the corresponding persistence functions of an $N$ spin chain. The correlation function of finite chains can then be obtained using Eq.~(\ref{eq:Paper2_CorrFunc}). In Fig.~\ref{fig:ShortChain}, we compare the 1-IF predictions with the exact solutions for finite chains of $N+1$ spins ($N=2,3,4$) and Monte Carlo simulations for a chain of $N+1=9$ spins ($N=8$), all at $c=0.01$.

The agreement for short chains indicates that the 1-IF approximation captures the dominant relaxation processes in small systems. As the chain length increases, however, the dynamics involves increasingly complex correlations between successive spin-flip intervals, which are neglected by the factorization approximation. The resulting accumulation of errors becomes particularly apparent in the long-time relaxation, where the 1-IF approximation predicts a faster decay than that observed in the simulations. This suggests that higher-order interval correlations become increasingly important as the system size grows.

\section{Conclusion and discussion}
\label{sec:discussion}
In this work, we develop a theoretical framework for the dynamics of the East model based on path probability densities. By exploiting the spatial translational invariance of the equilibrium dynamics, we derive a closed infinite system of equations for the marginal path probability densities of a single spin. 
Upon Laplace transformation, the original multiple-integral equations are converted into algebraic difference equations, providing a basis for the systematic construction of approximations.

We then introduce a systematic factorization approximation to solve this infinite system. At the $k$-th order, the approximation neglects memory effects associated with spin-flip histories beyond $k$ consecutive flips. This imposes an upper bound on the number of flips retained in the path-probability densities, thereby reducing the infinite system  to a finite set of equations. In this work, we focus on the first-order approximation, referred to as the 1-IF approximation.

Surprisingly, the results obtained using 1-IF agree remarkably well with those obtained using the 2-DB approximation in the Zwanzig–Mori formalism, both for the time correlation functions (see Fig.~\ref{fig:Paper2_CtCompareSIFVanZon}) and for the relaxation times (see Fig.~\ref{fig:TauCompareAll}). This close agreement raises the possibility that the two approaches capture similar aspects of the dynamics despite being based on different approximation schemes. In particular, truncating the evolution of the basis in the Zwanzig–Mori framework may suppress certain dynamical memory effects in a way that is related to the neglect of longer spin-flip histories in the path-probability framework. Establishing a precise connection between these two approximations, however, requires further investigation.

The present framework provides a formally closed description of the single-spin marginal path-probability densities, while the  factorization approximation offers a systematic route to obtaining finite-order approximations to this infinite system. Future work will focus on developing more effective methods for solving the closed equations and, in particular, on clarifying the relationship between the path-probability and Zwanzig–Mori approaches. Such an analysis may help establish a deeper understanding of the approximations underlying both frameworks.

Extending the present path-probability formalism to other models is an important future direction. Natural candidates include models with symmetric kinetic constraints, such as the Fredrickson--Andersen model in one dimension, and Hamiltonian models such as the Ising model in one dimension. It would then be very interesting to seek a generalization to two and three dimensions.

\section{Acknowledgement}
We thank Junru Wu, Chengjie Luo, Liesbeth Janssen, and Haijun Zhou for useful discussions. We acknowledge funding from the National Key R\&D Program of China (Grant No.~2025YFF0512000), the China Manned Space Program (Grant No.~CMSS-2025-5-P-002), the Wenzhou Institute (No.~WIUCASICTP2022), and the National Natural Science Foundation of China (Nos.~12447101).
We acknowledge the use of the High-Performance Cluster at the Institute of Theoretical Physics, Chinese Academy of Sciences.

\appendix

\section{Formulas used in the Laplace transform}
\label{app:first}

\subsection{Laplace transform of the multi-time integral}

As shown in Sec.~\ref{sec:kernel}, the kernel of the integral in Eq.~(\ref{eq:Paper2_FinalExplicitEqSystem}) has a simple structure: it contains a product of exponential functions. Since multiplication by an 
exponential shifts the Laplace variable, it is natural to analyze 
Eq.~(\ref{eq:Paper2_FinalExplicitEqSystem}) in the Laplace domain.

We define the Laplace transform of a multivariable function 
\(f(t_1,\dots,t_m)\) as
\begin{equation}
\begin{split}
   & \tilde{f}(z_1,\dots,z_m) \\
  & = \mathcal{L}_{z_1, \ldots, z_m} \{ f(t_1,\dots,t_m) \}
   \\
   & \equiv \int_0^{\infty} dt_1 e^{-z_1t_1}\cdots 
       \int_0^{\infty} dt_m e^{-z_mt_m}\,
       f(t_1,\dots,t_m),
\end{split}
    \label{eq:Paper2_DefOfLT}
\end{equation}
where \(\mathcal{L}_{z_1, \ldots, z_m}\{\cdots \}\) denotes the multidimensional Laplace transform. We use a tilde to denote the Laplace transform of a function.

\subsection{Generalized convolution theorem}

The generalized convolution theorem reads
\begin{equation}
\begin{split}
&\mathcal{L}_z\left\{
\int_0^\infty \left(\prod_{\alpha=1}^m dt'_\alpha\right)
\delta\!\left(t-\sum_{\alpha=1}^m t'_\alpha\right)
f(t'_1,\ldots,t'_m)
\right\}\\
&= \int_0^\infty dt\, e^{-zt}
 \left(\prod_{\alpha=1}^m dt'_\alpha\right)
\delta\!\left(t-\sum_{\alpha=1}^m t'_\alpha\right)
f(t'_1,\ldots,t'_m)\\
&= \tilde{f}(z_1,\ldots,z_m)\big|_{z_1=\cdots=z_m=z}\\
&= \tilde{f}(z\,\mathbf{1}_m),
\end{split}
\label{eq:convolution}
\end{equation}
where \(\mathbf{1}_m=(1,1,\ldots,1)\) is the all-ones vector of length \(m\).

Applying the generalized convolution theorem to
Eq.~(\ref{eq:Paper2_CorrToTrajectory}), we convert the 
multidimensional integral into an algebraic expression for the 
Laplace-transformed return probability:
\begin{equation}
\begin{split}
   &\mathcal{L}_z\{P[n_i(t)=1\mid n_i(0)=1]\} 
   = \sum_{m=0}^{\infty}
   \tilde{\Lambda}_{2m}^{1}(z\,\mathbf{1}_{2m+1}).
\end{split}
   \label{eq:Paper2_LTOfCorrToTraj}
\end{equation}

\subsection{Divided difference formula}

Consider a function $f(t)$ whose Laplace transform is $\tilde{f}(z)=\int_0^{\infty}e^{-zt}f(t)dt$. We need to express the Laplace transform of $f(t_1+t_2)$ with respect to two nodes $t_1,t_2$ in terms of $\tilde{f}(z)$:
\begin{equation}
    \mathcal{L}_{z_1,z_2}f(t_1+t_2)=\int_{0}^{\infty}e^{-z_1t_1}dt_1e^{-z_2t_2}dt_2f(t_1+t_2).
    \label{eq:LaplaceOfSumProblem}
\end{equation}
We make a change of variables
\begin{equation}
    \begin{cases}
        u=t_1+t_2,\\
        v=t_1-t_2.
    \end{cases}
    \label{eq:f_ChangeVariablesOf}
\end{equation}
Then, the inverse transform reads
\begin{equation}
    \begin{cases}
        t_1=\frac{u+v}{2},\\
        t_2=\frac{u-v}{2},
    \end{cases}
    \label{eq:Paper2_f_InverseChangeVar}
\end{equation}
followed by the transform of the integration region
\begin{equation}
    \begin{cases}
        0<t_1<t,\\
        0<t_2<t,
    \end{cases}\to\begin{cases}
        -u<v<-u+2t,\\
        u-2t<v<u.
    \end{cases}
    \label{eq:RegionChange}
\end{equation}

Substituting Eq.~(\ref{eq:f_ChangeVariablesOf}) and Eq.~(\ref{eq:RegionChange}) into Eq.~(\ref{eq:LaplaceOfSumProblem}), one can obtain
\begin{equation}
    \begin{split}
        &\mathcal{L}_{z_1,z_2}f(t_1+t_2)\\
        &=\lim_{t\to\infty}\frac{1}{2}\Bigg\{\left[\int_0^tdu\int_{-u}^{u}dv+\int_t^{2t}du\int_{u-2t}^{2t-u}dv\right]\\
        &\quad\quad\quad\quad\quad\quad\quad\quad\quad\quad\quad\quad\quad \times e^{-\frac{z_1+z_2}{2}u-\frac{z_1-z_2}{2}v}f(u)\Bigg\}\\
        &=\int_0^{\infty}\frac{e^{-z_1u}-e^{-z_2u}}{z_2-z_1}f(u)du\\
        &=-\frac{\tilde{f}(z_1)-\tilde{f}(z_2)}{z_1-z_2}\\
        &=-\mathcal{S}_{1,1}^{\{z_1,z_2\}}\tilde{f}(z').
    \end{split}
    \label{eq:LaplaceToDividedDiff}
\end{equation}
By mathematical induction, we can prove the following more general result:
\begin{equation}
\begin{split}
    \mathcal{L}_{z_1,z_2,...,z_m}f\left(\sum_{j=1}^{m}t_j \right)&=(-1)^{m-1}\mathcal{S}_{1,m-1}^{\{z_i\}}\tilde{f}(z')\\
    &=
(-1)^{m-1}
\sum_{j=1}^{m}
\frac{\tilde{f}(z_j)}
{\prod_{k\ne j}(z_j-z_k)}.
\end{split}
    \label{eq:nthDividDiffFormula}
\end{equation}

Equation~(\ref{eq:nthDividDiffFormula}) defines the divided-difference operator
\(\mathcal{S}_{\alpha,m-1}^{\{z_i\}}\). The subscript \(m-1\) denotes the order of
the divided difference. The subscript \(\alpha\) indicates that the operator acts
on the \(\alpha\)-th argument of \(\tilde{f}(\mathbf{z}')\); in the present case,
\(\mathbf{z}'\) is scalar and \(\alpha=1\). The superscript
\(\{z_i\}=\{z_1,z_2,\ldots,z_m\}\) specifies the nodes used for the divided
difference with respect to the \(\alpha\)-th argument.

\begin{figure}
    \centering
    \includegraphics[width=\linewidth]{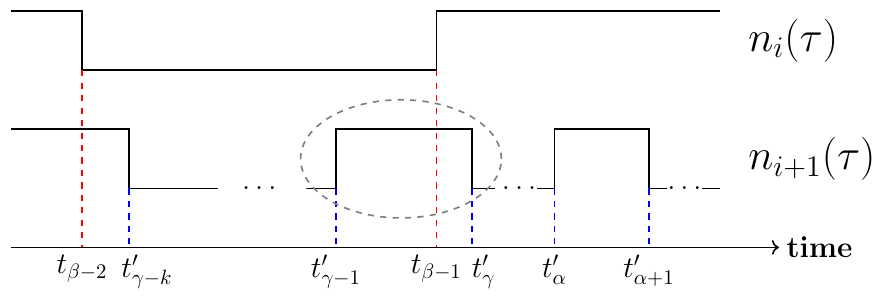}
    \caption{Schematic illustration of a $1$-plateau containing a single flip of spin $i$.
    }
    \label{fig:Paper2_TrajectoryoneFlip}
\end{figure}

\section{Kernel operator in the Laplace-domain}
\label{app:second}

We aim to obtain the Laplace-domain form of the path-probability equation, Eq.~(\ref{eq:Paper2_FinalExplicitEqSystem}). The Laplace transform of the left-hand side simply yields $\tilde{\Lambda}_m^{n_i}(\mathbf{z}_{m+1})$, where $\mathbf{z}_{m+1} = \{z_1, z_2, \ldots, z_{m+1}\}$. The Laplace transform of the right-hand side shows that the kernel $\mathcal{Q}^{n_i n_{i+1}}$ becomes an operator in the Laplace domain, which substantially changes how the Laplace transform acts on the integrals. Because of the exponential form of the kernel $\mathcal{Q}^{n_i n_{i+1}}$, these integrals can be expressed as sums involving shifted Laplace transforms and divided differences of the path-probability density functions appearing in the integrand. In this subsection, we analyze this transformation in detail.

As shown in  Fig.~\ref{fig:Paper2_PathOfTwoSpin}, the kernel has a form $(\omega_1 \omega_0 \ldots ) e^{-\omega_1(t_1 - t_1')} e^{-\omega_0(t'_2 - t_1) \ldots}$. We first derive the prefactor $(\omega_1 \omega_0 \ldots )$.  
In Eq.~(\ref{eq:Qab}), \(Q_{ab}^{0}=1\). Therefore, a 0-segment 
contributes a factor of one to the kernel and is unaffected by the 
Laplace transform. 
Each \(1\)-\(1\) segment contributes a 
factor of \(\omega_{n_i}\). By definition, a \(1\)-\(1\) segment ends with 
a flip of spin \(i\) (see Fig.~\ref{fig:Paper2_PathOfTwoSpin}). If spin 
\(i\) flips \(m\) times during the time interval \([0,t]\) of interest, 
then there are \(m\) such segments. Since the state of spin \(i\) 
alternates after each flip, the corresponding factors alternate between 
\(\omega_{n_i}\) and \(\omega_{\bar n_i}\), where 
\(\bar n_i=1-n_i\). Thus their total contribution to the kernel is
\(
\omega_{n_i}^{\lceil m/2\rceil}\omega_{\bar n_i}^{\lfloor m/2\rfloor},
\)
where \(\lceil \cdot \rceil\) and \(\lfloor \cdot \rfloor\) denote the 
ceiling and floor functions, respectively.
Once the prefactor is accounted for, the remaining part of \(Q_{ab}\) is a 
single exponential function for both \(1\)-\(1\) and \(1\)-\(0\) segments. 

Below, we derive the Laplace transform of the exponential part in the kernel. Recall that \(t_\alpha\) and \(t'_\alpha\) denote the flip times of spin \(i\) and spin \(i+1\), respectively. We have divided the total time interval into segments; within each segment, neither spin \(i\) nor spin \(i+1\) flips. We use \(i,j,\ldots\) to index the spins and \(\alpha,\beta,\ldots\) to index the flip times. Our strategy is as follows.

In general, one needs to consider two cases.

\begin{enumerate}
    \item[(i)] Spin \(i+1\) flips between two consecutive flips of spin \(i\) (see Fig.~\ref{fig:Paper2_Trajectory01NoFlipPlateau}). In this case, within the interval \(\Delta t_{\beta} = t_\beta - t_{\beta -1}\) between two flips of spin \(i\) at \(t_{\beta -1}\) and \(t_\beta\), there are several \(0\)- and \(1\)-\(0\) segments. The last segment is a \(1\)-\(1\) segment.

    \item[(ii)] Spin \(i\) flips one or multiple times between two consecutive flips of spin \(i+1\) (see Fig.~\ref{fig:Paper2_TrajectoryoneFlip} and Fig.~\ref{fig:Paper2_TrajectoryMultiFlip}). For spin \(i\) to flip, spin \(i+1\) must be in state \(1\) during the interval \(\Delta t'_{\alpha} = t'_\alpha - t'_{\alpha -1}\) between two flips at  \(t'_{\alpha-1}\) and \(t'_{\alpha}\). In this case, the interval \(\Delta t'_{\alpha}\) consists of a sequence of \(1\)-\(1\) segments followed by a final \(1\)-\(0\) segment. We call such a sequence \((1\text{-}1)\text{-}(1\text{-}1)\text{-}\cdots\text{-}(1\text{-}0)\) a \(1\)-plateau. Within each \(1\)-plateau, spin \(i+1\) remains in state \(1\), while spin \(i\) can flip one or multiple times.
\end{enumerate}

Having clarified these two cases, we then discuss the Laplace transform of the kernel over the entire time interval of interest \([0,t]\), which may contain multiple repetitions of the above two cases. A detailed analysis is provided below.

{\it (i) Spin \(i+1\) flips multiple times between two consecutive flips of spin \(i\).}\;
As shown in Fig.~\ref{fig:Paper2_Trajectory01NoFlipPlateau}, we consider two consecutive flips of spin $i$ at $t_{\beta-1}$ and $t_\beta$. Within the interval $\Delta t_{\beta}=t_{\beta}-t_{\beta-1}$, spin $i+1$ flips at the following times: $t'_{\gamma},t'_{\gamma+1},...,t'_{\gamma+l-1}$ where $l\geq 2$.  
Without loss of generality, we assume that spin $i$ remains in state $n_i$ throughout this interval.

Within $\Delta t_{\beta}$, each $0$-segment contributes a kernel factor $Q_{ab}^0(\Delta t'_\alpha)=1$, and each $1\text{-}0$ segment contributes $Q_{ab}^{1\text{-}0}(\Delta t'_{\alpha})=e^{-\omega_{n_i}\Delta t'_{\alpha}}$, where $\gamma+1\leq \alpha\leq \gamma+l-1$. For simplicity we redefine the time intervals of spin $i+1$ inside the $\beta$-th interval of spin $i$:
\begin{align}
    \begin{cases}
        \Delta t_1^{'\beta}=t'_{\gamma}-t_{\beta-1},\\
        \Delta t_\alpha^{'\beta}=\Delta t'_{\gamma+\alpha-1},\quad 2\leq\alpha\leq l,\\
        \Delta t_{l+1}^{'\beta}=t_{\beta}-t'_{\gamma+l-1}.
    \end{cases}
    \label{eq:Paper2_redefineBetaInterval}
\end{align}
Thus, these intervals satisfy the constraint:
\begin{equation}
    \sum_{\alpha=1}^{l+1}{\Delta t_{\alpha}^{'\beta}}=\Delta t_{\beta}.
    \label{eq:Paper2_PlateauInInterval_Constraint}
\end{equation}

Performing the Laplace transform with respect to $\Delta t_{\beta}$ and using the generalized convolution theorem Eq.~(\ref{eq:convolution}), we obtain
\begin{widetext}
    \begin{equation}
        \begin{split}
            &\mathcal{L}_{z_\beta}\Bigg\{\int_0^{\infty}\Bigg(\prod_{\alpha=1}^{l+1}d\Delta t_{\alpha}^{'\beta}Q_{ab}(\Delta t_{\alpha}^{'\beta})\Bigg)\delta\Big(\Delta t_\beta-\sum_{\alpha=1}^{l+1}\Delta t_{\alpha}^{'\beta}\Big)\Lambda_{m'}^{n_{i+1}}(\ldots,\Delta t'_{\gamma+1},\ldots,\Delta t'_{\gamma+l-1},\ldots)\Bigg\}\\
            &=\int_0^{\infty}e^{-z_\beta \Delta t_{\beta}}d\Delta t_{\beta}\Bigg(\prod_{\alpha=1}^{l+1}d\Delta t_{\alpha}^{'\beta}Q_{ab}(\Delta t_{\alpha}^{'\beta})\Bigg)\delta\Big(\Delta t_\beta-\sum_{\alpha=1}^{l+1}\Delta t_{\alpha}^{'\beta}\Big)\Lambda_{m'}^{n_{i+1}}(\ldots,\Delta t'_{\gamma+1},\ldots,\Delta t'_{\gamma+l-1},\ldots)\\
            &=\int_0^{\infty}\Bigg(\prod_{\alpha=1}^{l+1}d\Delta t_{\alpha}^{'\beta}e^{-z_{\beta}\Delta t_{\alpha}^{'\beta}}Q_{ab}(\Delta t_{\alpha}^{'\beta})\Bigg)\Lambda_{m'}^{n_{i+1}}(\ldots,\Delta t'_{\gamma+1},\ldots,\Delta t'_{\gamma+l-1},\ldots).
        \end{split}
    \label{eq:Paper2_LaplaceConvolution1_NoFlip}
    \end{equation}
\end{widetext}
The index of $1$-0 segments and $0$-segments runs from $\alpha=2$ to $\alpha=l$, so we focus on the contribution of these segments (see Fig.~\ref{fig:Paper2_Trajectory01NoFlipPlateau}).
The first and last segments ($\Delta t_1^{'\beta}$ and $\Delta t_{l+1}^{'\beta}$) contains one flip of spin $i$ -- this will be discussed in the next case. 

Substituting the kernel factors $Q_{ab}^0(\Delta t'_{\alpha})$ and $Q_{ab}^{(1\text{-}0)}(\Delta t'_{\alpha})$ in Eq.~(\ref{eq:Qab}) into Eq.~(\ref{eq:Paper2_LaplaceConvolution1_NoFlip}), we obtain
\begin{widetext}
    \begin{equation}
        \begin{split}
            &\int_0^{\infty}\Bigg(\prod_{\alpha=\gamma+1}^{\gamma+l-1}d\Delta t_{\alpha}'e^{-z_{\beta}\Delta t_{\alpha}'}Q_{ab}(\Delta t_{\alpha}')\Bigg)\Lambda_{m'}^{n_{i+1}}(\ldots,\Delta t'_{\gamma+1},\ldots,\Delta t'_{\gamma+l-1},\ldots)\\
            &=\int_0^{\infty}\Bigg(\prod_{\alpha}^{\{0\}}d\Delta t'_\alpha e^{-z_{\beta}\Delta t'_{\alpha}}\Bigg)\Bigg(\prod_{\alpha}^{\{1\text{-}0\}}d\Delta t'_{\alpha}e^{-(z_\beta+\omega_{n_i})\Delta t'_{\alpha}}\Bigg)\Lambda_{m'}^{n_{i+1}}(\ldots,\Delta t'_{\gamma+1},\ldots,\Delta t'_{\gamma+l-1},\ldots)\\
            &=\tilde{\Lambda}_{m'}^{n_{i+1}}(\ldots,\hat{z}_{\gamma+1},\ldots,\hat{z}_{\gamma+l-1},\ldots)\\
            &=\prod_{\alpha=\gamma+1}^{\gamma+l-1}\mathcal{S}_{\alpha,0}^{\hat{z}_{\alpha}}\tilde{\Lambda}_{m'}^{n_{i+1}}({\bf z}_{m'+1}'),
        \end{split}
    \label{eq:Paper2_LaplaceConvolution2_NoFlip}
    \end{equation}
\end{widetext}
where the products $\prod_{\alpha}^{\{0\}}$ and $\prod_{\alpha}^{\{1\text{-}0\}}$ run over the $0$-segments and $1\text{-}0$ segments, respectively. The Laplace variables associated with each $\Delta t'_{\alpha}$, with $\gamma+1\leq \alpha\leq \gamma+l-1$, is
\begin{equation}
    \hat{z}_{\alpha}=\begin{cases}
        z_\beta, & \text{for a $0$-segment},\\
        z_{\beta}+\omega_{n_i}, & \text{for a $1\text{-}0$ segment}.
    \end{cases}
    \label{eq:Paper2_hatz}
\end{equation}

Thus, in the present case, the kernel operator, denoted by $\mathcal{S}_{\alpha,0}^{\hat{z}_{\alpha}}$ in the last line of Eq.~(\ref{eq:Paper2_LaplaceConvolution2_NoFlip}), plays two roles for the $\alpha$-th interval of spin $i+1$. If this interval is a $0$-segment of spin $i+1$ within $\Delta t_{\beta}$ of spin $i$ whose state is $n_i$, $\mathcal{S}_{\alpha,0}^{\hat{z}_{\alpha}}$ replaces $z_{\alpha}'$ in $\tilde{\Lambda}_{m'}^{n_{i+1}}({\bf z}_{m'+1}')$ with $z_{\beta}$. If this interval is a $1\text{-}0$ segment of spin $i+1$ within $\Delta t_{\beta}$ of spin $i$, it replaces $z'_{\alpha}$ with the shifted value $z_{\beta}+\omega_{n_i}$, as shown in Eq.~(\ref{eq:Paper2_hatz}).

{\it (ii) $1$-plateau with one flip of spin $i$.}\;
In the previous case, we set aside the contributions from the two boundary $1$-plateaus containing the flip events of spin $i$ at $t_{\beta-1}$ and $t_{\beta}$, respectively. We now consider the contribution to the Laplace transform from the kernel associated with a $1$-plateau containing a single flip event of spin $i$.

Now we focus on the $1$-plateau corresponding to the interval $\Delta t'_{\gamma}=t'_{\gamma}-t'_{\gamma-1}$ in Fig.~\ref{fig:Paper2_Trajectory01NoFlipPlateau}. This $1$-plateau is the right boundary of the interval formed by two consecutive flips $t_{\beta-2},t_{\beta-1}$ of spin $i$, while it is the left boundary of that formed by another two consecutive flips $t_{\beta-1},t_{\beta}$. Hence the Laplace transform of the corresponding part of the integration is relevant to two convolution intervals. Within the interval $\Delta t_{\beta-1}=t_{\beta-1}-t_{\beta-2}$, spin $i+1$ flips at the following times: $t'_{\gamma-k},t'_{\gamma-k+1},...,t'_{\gamma-1}$ where $k\geq 2$. Similarly, we redefine the times intervals of spin $i+1$ inside the $\beta-1$-th interval of spin $i$ based on the same rule as Eq.~(\ref{eq:Paper2_redefineBetaInterval}):
\begin{align}
    \begin{cases}
        \Delta t_1^{'\beta-1}=t'_{\gamma-k}-t_{\beta-2},\\
        \Delta t_{\alpha}^{'\beta-1}=\Delta t'_{\gamma-k+\alpha-1},& 2\leq \alpha\leq k,\\
        \Delta t_{k+1}^{'\beta-1}=t_{\beta-1}-t'_{\gamma-1}.
    \end{cases}
    \label{eq:Paper2_redefineBetaMinors1Interval}
\end{align}
These intervals satisfy the constraint:
\begin{equation}
    \sum_{\alpha=1}^{k+1}\Delta t_{\alpha}^{'\beta-1}=\Delta t_{\beta-1}.
    \label{eq:Paper2_PlateauInInterval_betaMinors1_Constraint}
\end{equation}

Performing the Laplace transform with respect to $\Delta t_{\beta-1}$ and $\Delta t_{\beta}$, and using the generalized convolution theorem Eq.~(\ref{eq:convolution}), we obtain
\begin{widetext}
    \begin{equation}
        \begin{split}
            &\mathcal{L}_{z_{\beta-1},z_{\beta}}\Bigg\{\int_{0}^{\infty}\Bigg(\prod_{\alpha=1}^{k+1}d\Delta t_{\alpha}^{' \beta-1} Q_{ab}(\Delta t_{\alpha}^{' \beta-1})\Bigg)\Bigg(\prod_{\alpha'=1}^{l+1}d\Delta t_{\alpha'}^{' \beta} Q_{ab}(\Delta t_{\alpha'}^{' \beta})\Bigg)\\
            &\quad\quad\quad\quad\quad\quad\times\delta\Big(\Delta t_{\beta-1}-\sum_{\alpha=1}^{k+1}\Delta t_{\alpha}^{' \beta-1}\Big)\delta\Big(\Delta t_{\beta}-\sum_{\alpha'=1}^{l+1}\Delta t_{\alpha'}^{' \beta}\Big)\Lambda_{m'}^{n_{i+1}}(\ldots,\Delta t'_{\gamma},\ldots)\Bigg\}\\
            &=\int_0^{\infty}e^{-z_{\beta-1}\Delta t_{\beta-1}}d\Delta t_{\beta-1}e^{-z_{\beta}\Delta t_{\beta}}d\Delta t_{\beta}\Bigg(\prod_{\alpha=1}^{k+1}d\Delta t_{\alpha}^{' \beta-1} Q_{ab}(\Delta t_{\alpha}^{' \beta-1})\Bigg)\Bigg(\prod_{\alpha'=1}^{l+1}d\Delta t_{\alpha'}^{' \beta} Q_{ab}(\Delta t_{\alpha'}^{' \beta})\Bigg)\\
            &\quad\quad\quad\quad\quad\quad\times\delta\Big(\Delta t_{\beta-1}-\sum_{\alpha=1}^{k+1}\Delta t_{\alpha}^{' \beta-1}\Big)\delta\Big(\Delta t_{\beta}-\sum_{\alpha'=1}^{l+1}\Delta t_{\alpha'}^{' \beta}\Big)\Lambda_{m'}^{n_{i+1}}(\ldots,\Delta t'_{\gamma},\ldots)\\
            &=\int_0^{\infty}\Bigg(\prod_{\alpha=1}^{k+1}e^{-z_{\beta-1}\Delta t_{\alpha}^{' \beta-1}}d\Delta t_{\alpha}^{' \beta-1} Q_{ab}(\Delta t_{\alpha}^{' \beta-1})\Bigg)\Bigg(\prod_{\alpha'=1}^{l+1}e^{-z_{\beta}\Delta t_{\alpha'}^{' \beta}}d\Delta t_{\alpha'}^{' \beta} Q_{ab}(\Delta t_{\alpha'}^{' \beta})\Bigg)\Lambda_{m'}^{n_{i+1}}(\ldots,\Delta t'_{\gamma},\ldots).
        \end{split}
        \label{eq:Paper2_LaplaceConvolution1_oneFlip}
    \end{equation}
\end{widetext}
Equation~(\ref{eq:Paper2_LaplaceConvolution1_oneFlip}) connects the Laplace transform with respect to variables $\Delta t_{\beta-1}$ and $\Delta t_{\beta}$ to that with respect to $\{\Delta t'_\alpha\}$ within these two intervals of spin $i$. We now focus on the part contributed by interval $\Delta t'_{\gamma}=t'_{\gamma}-t'_{\gamma-1}$, because the other parts have been discussed in case (i) above. 
This reads
\begin{widetext}
    \begin{equation}
        \begin{split}
            &\int_0^{\infty}e^{-z_{\beta-1}\Delta t_{k+1}^{' \beta-1}}d\Delta t_{k+1}^{' \beta-1}e^{-z_{\beta}\Delta t_{1}^{' \beta}}d\Delta t_{1}^{' \beta}Q_{ab}^{1\text{-}1}(\Delta t_{k+1}^{' \beta-1})Q_{ab}^{1\text{-}0}(\Delta t_{1}^{' \beta})\Lambda_{m'}^{n_{i+1}}(\ldots,\Delta t_{k+1}^{' \beta-1}+\Delta t_{1}^{' \beta},\ldots)\\
            &=\int_0^{\infty}e^{-z_{\beta-1}\Delta t_{k+1}^{' \beta-1}}d\Delta t_{k+1}^{' \beta-1}e^{-z_{\beta}\Delta t_{1}^{' \beta}}d\Delta t_{1}^{' \beta}\omega_{\bar{n}_i}e^{-\omega_{\bar{n}_i}\Delta t_{k+1}^{' \beta-1}}e^{-\omega_{n_i}\Delta t_{1}^{' \beta}}\Lambda_{m'}^{n_{i+1}}(\ldots,\Delta t_{k+1}^{' \beta-1}+\Delta t_{1}^{' \beta},\ldots)\\
            &=\omega_{\bar{n}_i}\int_0^{\infty}e^{-(z_{\beta-1}+\omega_{\bar{n}_i})\Delta t_{k+1}^{' \beta-1}}d\Delta t_{k+1}^{' \beta-1}e^{-(z_{\beta}+\omega_{n_i})\Delta t_{1}^{' \beta}}d\Delta t_{1}^{' \beta}\Lambda_{m'}^{n_{i+1}}(\ldots,\Delta t_{k+1}^{' \beta-1}+\Delta t_{1}^{' \beta},\ldots)\\
            &=-\omega_{\bar{n}_i}\frac{\tilde{\Lambda}_{m'}^{n_{i+1}}(\ldots,z_{\gamma}',\ldots)|_{z_{\gamma}'=z_{\beta-1}+\omega_{\bar{n}_i}}-\tilde{\Lambda}_{m'}^{n_{i+1}}(\ldots,z_{\gamma}',\ldots)|_{z_{\gamma}'=z_{\beta}+\omega_{n_i}}}{z_{\beta-1}+\omega_{\bar{n}_i}-(z_{\beta}+\omega_{n_i})}\\
            &=-\omega_{\bar{n}_i}\mathcal{S}_{\gamma,1}^{\{\hat{z}_{\gamma}^{(1)},\hat{z}_{\gamma}^{(2)}\}}\tilde{\Lambda}_{m'}^{n_{i+1}}({\bf z}_{m'+1}').
        \end{split}
        \label{eq:Paper2_LaplaceConvolution2_oneFlip}
    \end{equation}
\end{widetext}
Here, we have used $\Delta t'_{\gamma}=\Delta t_{k+1}^{' \beta-1}+\Delta t_{1}^{' \beta}$. 

The kernel operator, denoted by $\mathcal{S}_{\gamma,1}^{\{\hat{z}^{(1)}_{\gamma},\hat{z}^{(2)}_{\gamma}\}}$ in the last line of Eq.~(\ref{eq:Paper2_LaplaceConvolution2_oneFlip}), is a divided difference operator defined in Eq.~(\ref{eq:nthDividDiffFormula}).
The subscript $1$ denotes a first-order divided difference, and the subscript $\gamma$ indicates that the divided difference is taken with respect to the $\gamma$-th argument of the function $\tilde{\Lambda}_{m'}^{n_{i+1}}({\bf z}_{m'+1}')$. The superscripts $\hat{z}_{\gamma}^{(1)}$ and $\hat{z}_{\gamma}^{(2)}$ specify two nodes of this divided difference, namely
\begin{equation}
    \hat{z}_{\gamma}^{(1)}=z_{\beta-1}+\omega_{\bar{n}_i},\hat{z}_{\gamma}^{(2)}=z_{\beta}+\omega_{n_i}.
    \label{eq:Paper2_hatz1stDivideDiff}
\end{equation}

{\it (iii) $1$-plateau with multiple flips of spin $i$.}\;
A $1$-plateau of spin $i+1$ may contain more than one flip of spin $i$. As shown in Fig.~\ref{fig:Paper2_TrajectoryMultiFlip}, we consider the $1$-plateau corresponding to the interval $\Delta t'_{\alpha}=t'_{\alpha}-t'_{\alpha-1}$. Within the interval $\Delta t'_{\alpha}$, spin $i$ flips at the following times: $t_{\beta},t_{\beta+1},...,t_{\beta + m -1}$ where $m\geq 2$. Without loss of generality, we assume that spin $i$ starts from state $n_i$ at the beginning of this one plateau. This $1$-plateau consists of a sequence of $1\text{-}1$ segments followed by a $1\text{-}0$ segment. We denote
\begin{align}
    \begin{cases}
        \Delta t_{1}'^{\alpha}=t_{\beta}-t'_{\alpha-1},\\
        \Delta t_{\gamma}'^{\alpha}=\Delta t_{\beta+\gamma-1}, & 2\leq\gamma\leq m,\\
        \Delta t_{n+1}'^{\alpha}=t'_{\alpha}-t_{\beta+m-1}
    \end{cases}
    \label{eq:Paper2_NewGapNotationMultiFlip}
\end{align}
for the time intervals. Each $1\text{-}1$ segment contributes a kernel factor 
\begin{equation}
    Q_{ab}^{1\text{-}1}(\Delta t_{\gamma}'^{\alpha})=\begin{cases}
        \omega_{n_i}e^{-\omega_{n_i}\Delta t_{\gamma}'^{\alpha}},& \gamma\text{ is odd},\\
        \omega_{\bar{n}_i}e^{-\omega_{\bar{n}_i}\Delta t_{\gamma}'^{\alpha}}, & \gamma\text{ is even,}
    \end{cases}
    \label{eq:Paper2_11KernelinMultiFlip}
\end{equation}
where $1\leq\gamma\leq m$.
The final $1\text{-}0$ segment contributes 
\begin{equation}
    Q_{ab}^{1\text{-}0}(\Delta t_{m+1}'^{\alpha})=\begin{cases}
        e^{-\omega_{n_i}\Delta t_{m+1}'^{\alpha}},& m+1\text{ is odd},\\
        e^{-\omega_{\bar{n}_i}\Delta t_{m+1}'^{\alpha}}, & m+1\text{ is even.}
    \end{cases}
    \label{eq:Paper2_10KernelinMultiFlip}
\end{equation}
As in the previous two cases, taking the Laplace transform with respect to variables $\Delta t_{\beta},\Delta t_{\beta+1},\ldots, \Delta t_{\beta +m}$ allows us to relate the integral with respect to these variables to the integral with respect to variables $\Delta t_{1}'^{\alpha},...,\Delta t_{m+1}'^{\alpha}$ via the convolution formula (\ref{eq:convolution}). Here, we focus on the contribution of this $1$-plateau, which can be written as:
\begin{widetext}
    \begin{equation}
        \begin{split}
            &\int_0^{\infty}\Bigg(\prod_{\gamma=1}^{m+1}Q_{ab}(\Delta t_{\gamma}'^{\alpha})e^{-z_{\beta+\gamma-1}\Delta t_{\gamma}'^{\alpha}}d\Delta t_{\gamma}'^\alpha\Bigg)\Lambda_{m'}^{n_{i+1}}(\ldots,\Delta t'_{\alpha},\ldots)\\
            &=\omega_{n_i}^{\lceil m/2\rceil}\omega_{\bar{n}_i}^{\lfloor m/2\rfloor}\int_0^{\infty}\Bigg(\prod_{\gamma}^{\text{odd}}e^{-(z_{\beta+\gamma-1}+\omega_{n_i})\Delta t_{\gamma}'^{\alpha}}\Bigg)\Bigg(\prod_{\gamma}^{\text{even}}e^{-(z_{\beta+\gamma-1}+\omega_{\bar{n}_i})\Delta t_{\gamma}'^{\alpha}}\Bigg)\Lambda_{m'}^{n_{i+1}}\left(\ldots,\sum_{\gamma=1}^{m+1}\Delta t_{\gamma}'^{\alpha},\ldots \right)\\
            &=(-1)^{m}\omega_{n_i}^{\lceil m/2\rceil}\omega_{\bar{n}_i}^{\lfloor m/2\rfloor}\, \mathcal{S}_{\alpha,m}^{\{\hat{z}_{\alpha}^{(\gamma)}\}} \, \tilde{\Lambda}_{m'}^{n_{i+1}}({\bf z}'_{m'+1}),
        \end{split}
        \label{eq:Paper2_LaplaceConvolution1_multipleFlip}
    \end{equation}
\end{widetext}
where $\prod_{\gamma}^{\text{odd}}$ and $\prod_{\gamma}^{\text{even}}$ run over all odd $\gamma$ and even $\gamma$, respectively. Here, we have used  $\Delta t'_{\alpha}=\sum_{\gamma=1}^{m+1}\Delta t_{\gamma}'^{\alpha}$.

In the present case, the kernel operator is denoted by $\mathcal{S}_{\alpha,m}^{\{\hat{z}^{(\gamma)}_{\alpha} \}}$ in the last line of Eq.~(\ref{eq:Paper2_LaplaceConvolution1_multipleFlip}), which is the generalization of operator $\mathcal{S}_{\gamma,1}^{\{\hat{z}_{\gamma}^{(1)},\hat{z}_{\gamma}^{(2)}\}}$ in Eq.~(\ref{eq:Paper2_LaplaceConvolution2_oneFlip}). The subscript $m$ in  $\mathcal{S}_{\alpha,m}^{\{\hat{z}^{(\gamma)}_{\alpha} \}}$ denotes a $m$-th-order divided difference, and the subscript $\alpha$ indicates that the divided difference is taken with respect to the $\alpha$-th argument of the function $\tilde{\Lambda}_{m'}^{n_{i+1}}({\bf z}'_{m'+1})$. The superscript $\{\hat{z}_{\alpha}^{(\gamma)} \}$ with $
1\leq \gamma\leq m+1$ specifies the $m+1$ nodes of the divided difference, namely
\begin{equation}
    \hat{z}_{\alpha}^{(\gamma)}=\begin{cases}
        z_{\beta+\gamma-1}+\omega_{n_i}, & \gamma\text{ is odd},\\
        z_{\beta+\gamma-1}+\omega_{\bar{n}_i}, & \gamma\text{ is even}.
        \label{eq:Paper2_hatzMultiFlipA}
    \end{cases}
\end{equation}

\section{Two-interval factorization approximation}
\label{sec:two-interval}

Following the idea of 1-IF, we extend the scheme to second order, i.e., 2-IF. We assume that decorrelation occurs after every two consecutive flips of spin \(i\). Under this assumption, the Laplace-transformed path-probability densities \(\tilde{\Lambda}_m^{n_i}(\mathbf{z}_{m+1})\) can be factorized as follows:
\begin{align}
    \begin{split}
        &\tilde{\Lambda}_{2m-1}^{n_i}({\bf z}_{2m})=\\
        &\quad\quad\tilde{F}_1^{n_i}(z_1,z_2)\prod_{\alpha=2}^{m-1}\Big[\tilde{G}_1^{n_i}(z_{2\alpha-1},z_{2\alpha})\Big]\tilde{\Xi}_1^{n_i}(z_{2m-1},z_{2m}),\\
        &\tilde{\Lambda}_{2m}^{n_i}({\bf z}_{2m+1})=\\
        &\quad\quad\tilde{F}_1^{n_i}(z_1,z_2)\prod_{\alpha=2}^{m}\Big[\tilde{G}_1^{n_i}(z_{2\alpha-1},z_{2\alpha})\Big]\tilde{\Xi}_0^{n_i}(z_{2m+1}),
    \end{split}
    \label{eq:Paper2_Decomposition2ndOrder}
\end{align}
where $\tilde{F}_1^{n_i}(z_1,z_2)$ is the Laplace transform of $F_1^{n_i}(\Delta t_1,\Delta t_2)$ associated with the first segment. It denotes the next-flip-time density associated with $\Lambda_1^{n_i}(\Delta t_1,\Delta t_2)$.

The relationship between $F_1^{n_i}$ and $\Lambda_1^{n_i}$ can be derived as follows. The path-probability density $\Lambda_1^{n_i}(\Delta t_1,\Delta t_2)$ describes trajectories in which spin $i$ undergoes zero flips during $\Delta t_2=t-t_1$, and the second flip can occur at any time after $t$. Thus, $\Lambda_1^{n_i}(\Delta t_1,\Delta t_2)$ can be obtained by integrating the next-flip-time density $F_1^{n_i}(\Delta t_1,\Delta \tau)$ over all possible second-flip times $\Delta \tau>\Delta t_2$, yielding
\begin{equation}
    \Lambda_1^{n_i}(\Delta t_1,\Delta t_2)=\int_{\Delta t_2}^{\infty}d\Delta \tau F_1^{n_i}(\Delta t_1,\Delta \tau).
    \label{eq:Paper2_RelationOfPath&NextFlipDensityF1}
\end{equation}

For the intermediate segments, the probabilities are described by $G_{1}^{n_i}(\Delta t_{\alpha},\Delta t_{\alpha+1})$, which is the conditional next-flip-time density associated with $\Xi_1^{n_i}(\Delta t_{\alpha},\Delta t_{\alpha+1})$. We have
\begin{equation}
    \Xi_1^{n_i}(\Delta t_1,\Delta t_2)=\int_{\Delta t_2}^{\infty}d\Delta \tau G_1^{n_i}(\Delta t_1,\Delta \tau).
    \label{eq:Paper2_Xi1G1}
\end{equation}

Performing Laplace transform on Eqs.~\eqref{eq:Paper2_RelationOfPath&NextFlipDensityF1} and \eqref{eq:Paper2_Xi1G1}, we obtain
\begin{equation}
    \begin{split}
        &\tilde{F}_1^{n_i}(z_1,z_2)=\tilde{F}_1^{n_i}(z_1)-z_2\tilde{\Lambda}_1^{n_i}(z_1,z_2),\\
        &\tilde{G}_1^{n_i}(z_1,z_2)=\tilde{G}_1^{n_i}(z_1)-z_2\tilde{\Xi}_1^{n_i}(z_1,z_2).
    \end{split}
    \label{eq:Paper2_LaplaceOfF1Lambda1G1Xi1}
\end{equation}

Setting $m=1$ in Eq.~\eqref{eq:Paper2_FinalLaplaceExpression}, we obtain
\begin{widetext}
    \begin{equation}
        \begin{split}
            \tilde{\Lambda}_1^{n_i}(z_1,z_2)&=-\omega_{n_i}\sum_{m'=0}^{\infty}\sum_{\{x_\alpha\}}'\prod_{\alpha=1}^{m'+1}\mathcal{S}_{\alpha,x_{\alpha}}^{\{\hat{z}_{\alpha}^{(\gamma_\alpha)}\}}\Bigg[c\tilde{\Lambda}_{m'}^{1}({\bf z}'_{m'+1})+(1-c)\tilde{\Lambda}_{m'}^{0}({\bf z}'_{m'+1})\Bigg]\\
            &=-\omega_{n_i}\sum_{m'=0}^{\infty}\sum_{\beta=1}^{m'+1}\mathcal{S}_{\beta,1}^{\{\hat{z}_{\beta}^{(1)},\hat{z}_{\beta}^{(2)}\}}\prod_{\alpha\neq \beta}\mathcal{S}_{\alpha,0}^{\hat{z}_{\alpha}}\Bigg[c\tilde{\Lambda}_{m'}^{1}({\bf z}'_{m'+1})+(1-c)\tilde{\Lambda}_{m'}^{0}({\bf z}'_{m'+1})\Bigg].
        \end{split}
        \label{eq:Paper2_FinalLaplace_m=1}
    \end{equation}
\end{widetext}

Eq.~\eqref{eq:Paper2_FinalLaplace_m=1} relates $\tilde{\Lambda}_1^{n_i}(z_1,z_2)$ of spin $i$ to $\tilde{\Lambda}_{m'}^{0}({\bf z}_{m'+1})$ and $\tilde{\Lambda}_{m'}^{1}({\bf z}_{m'+1})$ of spin $i+1$. Similarly, we can relate $\tilde{\Xi}_1^{n_i}(z_1,z_2)$ to probabilities of spin $i+1$. Since $\tilde{\Xi}_1^{n_i}(z_1,z_2)$ is the path-probability density for trajectories in which spin $i+1$ is initially in state $1$, it is related only to $\tilde{\Lambda}_{m'}^{1}({\bf z}_{m'+1})$, as:
    \begin{equation}
        \begin{split}
            &\tilde{\Xi}_1^{n_i}(z_1,z_2)\\
            &=-\omega_{n_i}\sum_{m'=0}^{\infty}\sum_{\beta=1}^{m'+1}\mathcal{S}_{\beta,1}^{\{\hat{z}_{\beta}^{(1)},\hat{z}_{\beta}^{(2)}\}}\prod_{\alpha\neq \beta}\mathcal{S}_{\alpha,0}^{\hat{z}_{\alpha}}\tilde{\Lambda}_{m'}^{1}({\bf z}'_{m'+1}).
        \end{split}
        \label{eq:Paper2_FinalLaplace_m=1_Xi}
    \end{equation}
Substituting Eq.~\eqref{eq:Paper2_Decomposition2ndOrder} into Eq.~\eqref{eq:Paper2_FinalLaplace_m=0}, Eq.~\eqref{eq:Paper2_FinalLaplace_m=0_Xi}, Eq.~\eqref{eq:Paper2_FinalLaplace_m=1} and Eq.~\eqref{eq:Paper2_FinalLaplace_m=1_Xi}, and using Eq.~\eqref{eq:Paper2_LaplaceOfF1Lambda1G1Xi1}, we obtain the closed equation at the second order:
\begin{widetext}
    \begin{equation}
        \begin{split}
            &\tilde{\Lambda}_0^{n_i}(z)=(1-c)\frac{\tilde{\Xi}_0^0(z)+\tilde{\Xi}_0^1(z+\omega_{n_i})-\omega_{n_i}\tilde{\Xi}_0^0(z)\tilde{\Lambda}_1^0(z,z+\omega_{n_i})-z\tilde{\Lambda}_0^0(z)\big[\tilde{\Xi}_0^0(z)+\tilde{\Xi}_1^0(z,z+\omega_{n_i})\big]}{z\tilde{\Xi}_0^0(z)+(z+\omega_{n_i})\tilde{\Xi}_1^0(z,z+\omega_{n_i})}\\
            &\quad\quad\quad+c\frac{\tilde{\Xi}_0^1(z+\omega_{n_i})+\tilde{\Xi}_1^1(z+\omega_{n_i},z)+\omega_{n_i}\big[\tilde{\Lambda}_1^1(z+\omega_{n_i},z)\tilde{\Xi}_0^1(z+\omega_{n_i})-\tilde{\Lambda}_0^1(z+\omega_{n_i})\tilde{\Xi}_1^1(z+\omega_{n_i},z)\big]}{(z+\omega_{n_i})\tilde{\Xi}_0^1(z+\omega_{n_i})+z\tilde{\Xi}_1^0(z,z+\omega_{n_i})},\\
            &\tilde{\Xi}_{0}^{n_i}(z)=\frac{\tilde{\Xi}_0^1(z+\omega_{n_i})+\tilde{\Xi}_1^1(z+\omega_{n_i},z)+\omega_{n_i}\big[\tilde{\Lambda}_1^1(z+\omega_{n_i},z)\tilde{\Xi}_0^1(z+\omega_{n_i})-\tilde{\Lambda}_0^1(z+\omega_{n_i})\tilde{\Xi}_1^1(z+\omega_{n_i},z)\big]}{(z+\omega_{n_i})\tilde{\Xi}_0^1(z+\omega_{n_i})+z\tilde{\Xi}_1^0(z,z+\omega_{n_i})},\\
            &\tilde{\Lambda}_1^{n_i}(z_1,z_2)=-\omega_{n_i}\Big[(1-c)(\mathcal{A}_1+\mathcal{A}_2)+c(\mathcal{B}_1+\mathcal{B}_2)\Big],\\
            &\tilde{\Xi}_1^{n_i}(z_1,z_2)=-\omega_{n_i}(\mathcal{B}_1+\mathcal{B}_2).
        \end{split}
        \label{eq:Paper2_2ndCloseRecursion}
    \end{equation}
\end{widetext}
Here,
\begin{widetext}
    \begin{equation}
        \begin{split}
            &\mathcal{A}_1=\frac{\tilde{\Lambda}_1^0(z_1,\hat{z}_1)-\tilde{\Lambda}_1^0(z_1,\hat{z}_2)}{\hat{\Delta}_{12}}+\frac{[\hat{z}_2\tilde{\Lambda}_1^0(z_1,\hat{z}_2)-\hat{z}_1\tilde{\Lambda}_1^0(z_1,\hat{z}_1)]\tilde{\Xi}_1^0(z_2,\hat{z}_2)}{\hat{\Delta}_{12}[z_2\tilde{\Xi}_0^0(z_2)+\hat{z}_2\tilde{\Xi}_1^0(z_2,\hat{z}_2)]}\\
            &\quad\quad+\frac{[1-z_1\tilde{\Lambda}_0^0(z_1)-\hat{z}_1\tilde{\Lambda}_1^0(z_1,\hat{z}_1)][\hat{z}_2\tilde{\Xi}_1^0(z_1,\hat{z}_2)-\hat{z}_1\tilde{\Xi}_1^0(z_1,\hat{z}_1)]\tilde{\Xi}_1^0(z_2,\hat{z}_2)}{\hat{\Delta}_{12}[z_1\tilde{\Xi}_0^0(z_1)+\hat{z}_1\tilde{\Xi}_1^0(z_1,\hat{z}_1)][z_2\tilde{\Xi}_0^0(z_2)+\hat{z}_2\tilde{\Xi}_1^0(z_2,\hat{z}_2)]}\\
            &\quad\quad+\frac{[1-z_1\tilde{\Lambda}_0^0(z_1)-\hat{z}_1\tilde{\Lambda}_1^0(z_1,\hat{z}_1)][\tilde{\Xi}_1^0(z_1,\hat{z}_1)-\tilde{\Xi}_1^0(z_1,\hat{z}_2)]}{\hat{\Delta}_{12}[z_1\tilde{\Xi}_0^0(z_1)+\hat{z}_1\tilde{\Xi}_1^0(z_1,\hat{z}_1)]},\\
            &\mathcal{A}_2=\frac{[\hat{z}_2\tilde{\Lambda}_1^0(z_1,\hat{z}_2)-\hat{s}_1\tilde{\Lambda}_1^0(z_1,\hat{z}_1)]\tilde{\Xi}_0^0(z_2)}{\hat{\Delta}_{12}[z_2\tilde{\Xi}_0^0(z_2)+\hat{z}_2\tilde{\Xi}_1^0(z_2,\hat{z}_2)]}+\frac{[1-z_1\tilde{\Lambda}_0^0(z_1)-\hat{z}_1\tilde{\Lambda}_1^0(z_1,\hat{z}_1)][\hat{z}_2\tilde{\Xi}_1^0(z_1,\hat{z}_2)-\hat{z}_1\tilde{\Xi}_1^0(z_1,\hat{z}_1)]\tilde{\Xi}_0^0(z_2)}{\hat{\Delta}_{12}[z_1\tilde{\Xi}_0^0(z_1)+\hat{z}_1\tilde{\Xi}_1^0(z_1,\hat{z}_1)][z_2\tilde{\Xi}_0^0(z_2)+\hat{z}_2\tilde{\Xi}_1^0(z_2,\hat{z}_2)]},\\
&\mathcal{B}_1=\frac{\tilde{\Lambda}_0^1(\hat{z}_1)-\tilde{\Lambda}_0^1(\hat{z}_2)}{\hat{\Delta}_{12}}+\frac{\{\hat{z}_2\tilde{\Lambda}_0^1(\hat{z}_2)-\hat{z}_1\tilde{\Lambda}_0^1(\hat{z}_1)+z_2[\tilde{\Lambda}_1^1(\hat{z}_2,z_2)-\tilde{\Lambda}_1^1(\hat{z}_1,z_2)]\}\tilde{\Xi}_0^1(\hat{z}_2)}{\hat{\Delta}_{12}[\hat{z}_2\tilde{\Xi}_0^1(\hat{z}_2)+{z}_2\tilde{\Xi}_1^1(\hat{z}_2,{z}_2)]}\\
            &\quad\quad+\frac{[1-\hat{z}_1\tilde{\Lambda}_0^1(\hat{z}_1)-z_1\tilde{\Lambda}_1^1(\hat{z}_1,z_1)]\{\hat{z}_2\tilde{\Xi}_0^1(\hat{z}_2)-\hat{z}_1\tilde{\Xi}_0^1(\hat{z}_1)+z_2[\tilde{\Xi}_1^1(\hat{z}_2,z_2)-\tilde{\Xi}_1^1(\hat{z}_1,z_2)]\}\tilde{\Xi}_0^1(\hat{z}_2)}{\hat{\Delta}_{12}[\hat{z}_1\tilde{\Xi}_0^1(\hat{z}_1)+{z}_1\tilde{\Xi}_1^1(\hat{z}_1,{z}_1)][\hat{z}_2\tilde{\Xi}_0^1(\hat{z}_2)+{z}_2\tilde{\Xi}_1^1(\hat{z}_2,{z}_2)]}\\
            &\quad\quad+\frac{[1-\hat{z}_1\tilde{\Lambda}_0^1(\hat{z}_1)-z_1\tilde{\Lambda}_1^1(\hat{z}_1,z_1)][\tilde{\Xi}_0^1(\hat{z}_1)-\tilde{\Xi}_0^1(\hat{z}_2)]}{\hat{\Delta}_{12}[\hat{z}_1\tilde{\Xi}_0^1(\hat{z}_1)+{z}_1\tilde{\Xi}_1^1(\hat{z}_1,{z}_1)]},\\
            &\mathcal{B}_2=\frac{\tilde{\Lambda}_1^1(\hat{z}_1,z_2)-\tilde{\Lambda}_1^1(\hat{z}_2,z_2)}{\hat{\Delta}_{12}}+\frac{\{\hat{z}_2\tilde{\Lambda}_0^1(\hat{z}_2)-\hat{z}_1\tilde{\Lambda}_0^1(\hat{z}_1)+z_2[\tilde{\Lambda}_1^1(\hat{z}_2,z_2)-\tilde{\Lambda}_1^1(\hat{z}_1,z_2)]\}\tilde{\Xi}_1^1(\hat{z}_2,{z}_2)}{\hat{\Delta}_{12}[\hat{z}_2\tilde{\Xi}_0^1(\hat{z}_2)+{z}_2\tilde{\Xi}_1^1(\hat{z}_2,{z}_2)]}\\
            &\quad\quad+\frac{[1-\hat{z}_1\tilde{\Lambda}_0^1(\hat{z}_1)-z_1\tilde{\Lambda}_1^1(\hat{z}_1,z_1)]\{\hat{z}_2\tilde{\Xi}_0^1(\hat{z}_2)-\hat{z}_1\tilde{\Xi}_0^1(\hat{z}_1)+z_2[\tilde{\Xi}_1^1(\hat{z}_2,z_2)-\tilde{\Xi}_1^1(\hat{z}_1,z_2)]\}\tilde{\Xi}_1^1(\hat{z}_2,z_2)}{\hat{\Delta}_{12}[\hat{z}_1\tilde{\Xi}_0^1(\hat{z}_1)+{z}_1\tilde{\Xi}_1^1(\hat{z}_1,{z}_1)][\hat{z}_2\tilde{\Xi}_0^1(\hat{z}_2)+{z}_2\tilde{\Xi}_1^1(\hat{z}_2,{z}_2)]}\\
            &\quad\quad+\frac{[1-\hat{z}_1\tilde{\Lambda}_0^1(\hat{z}_1)-z_1\tilde{\Lambda}_1^1(\hat{z}_1,z_1)][\tilde{\Xi}_1^1(\hat{z}_1,z_2)-\tilde{\Xi}_1^1(\hat{z}_2,z_2)]}{\hat{\Delta}_{12}[\hat{z}_1\tilde{\Xi}_0^1(\hat{z}_1)+{z}_1\tilde{\Xi}_1^1(\hat{z}_1,{z}_1)]},
        \end{split}
        \label{eq:Paper2_B1B2}
    \end{equation}
\end{widetext}
where $\hat{z}_1=z_1+\omega_{n_i},\hat{z}_2=z_2+\omega_{\bar{n}_i}$ and $\hat{\Delta}_{12}=\hat{z}_1-\hat{z}_2$.

\section{Comparison between single-interval factorization and two-domain basis theories}
\label{eq:comp_SIF_2domain}

Our strategy is as follows. We aim to compare the four independent persistence functions $\tilde{\Lambda}_0^1(z)$, $\tilde{\Lambda}_0^0(z)$, $\tilde{\Xi}_1^1(z)$, and $\tilde{\Xi}_1^1(z)$, obtained from the 1-IF equations, Eqs.~\eqref{eq:Paper2_1stCloseRecusion1} and~\eqref{eq:Paper2_1stCloseRecusion2}, with those predicted by the 2-DB theory. To this end, we first derive four equations that relate the (two-point) correlation functions $\tilde{C}(z)$ and the three-point correlation functions $\tilde{C}_{1,2}(z)$ to the four persistence functions. Solving these equations then allows us to express the persistence functions in terms of $\tilde{C}(z)$ and $\tilde{C}_{1,2}(z)$, whose expressions are given by the 2-DB theory~\cite{van2005glassy}.
We then compare the persistence functions from the two theories numerically and show that they are in close agreement.

\subsection{Correlation function in terms of persistence functions}

As shown in Eq.~\eqref{eq:Paper2_CorrFunc}, the correlation function can be expressed in terms of the return probability. In fact, a more general relationship exists between equilibrium time correlation functions and the time-dependent mean occupation under a conditionally prepared initial state~\cite{mauch1999recursive}:
\begin{equation}
    C(t)=\frac{\langle n_i\rangle_{t}-c}{\langle n_i\rangle_{t=0}-c}.
    \label{eq:Paper2_CorrToNonEq}
\end{equation}
To construct this initial state, the spin at site $i$ is assigned an arbitrary initial distribution $P_{t=0}(n_i)$ that can differ from the equilibrium distribution, while all spins to the right of $i$ are drawn from the equilibrium distribution. The averages $\langle \ldots \rangle_t$ are taken over trajectories evolving from this conditionally prepared initial ensemble.

To prove Eq.~(\ref{eq:Paper2_CorrToNonEq}), let us assume that the initial distribution of spin $i$ can be written in a perturbative form,
\begin{equation}
    P_{t=0}(n_i)=P_{\rm eq}(n_i)\Delta_1(n_i).
    \label{eq:Paper2_niInitial}
\end{equation}
Thanks to the property $n_i^2=n_i$, any higher-order term reduces to a linear term $n_i^2=n_i^3 = \ldots = n_i$. Thus,
a general expansion of  
the deviation of the distribution becomes linear,
\begin{equation}
    \Delta_1(n_i)=1+a\delta n_{i}.
    \label{eq:Paper2_deviateni}
\end{equation}
The initial average of $n_i$ is:
\begin{equation}
    \langle n_i\rangle_{t=0}=c+ac(1-c).
    \label{eq:Paper2_Initialni}
\end{equation}
Expressing $a$ in terms of $\langle n_i\rangle_{t=0}$, we obtain
\begin{equation}
    P_{t=0}(n_i)=P_{\rm eq}(n_i)\left[1+\frac{\langle \delta n_i\rangle_{t=0}}{c(1-c)}\delta n_i\right].
    \label{eq:Paper2_niFinalInitial}
\end{equation}

The stochastic dynamics of the East model are governed by the master equation Eq.~(\ref{eq:masterEQ}). Its solution can be expressed using the time-evolution operator:
\begin{equation}
    P({\bf n},t)=e^{-\mathcal{L}t}P({\bf n},t=0).
    \label{eq:Paper2_timeevolOfPnt}
\end{equation}
Since all spins to the right of spin $i$ are initially sampled from equilibrium, the time-evolution operator acting on their initial distribution yields unity. After tracing out the spins to the right of $i$, the marginal distribution of $n_i$ evolves as
\begin{equation}
    \begin{split}
         P_t(n_i)&=e^{-\mathcal{L}t}P_{t=0}(n_i)\\
         &=P_{\rm eq}(n_i)+\frac{\langle\delta n_i\rangle_{t=0}}{c(1-c)}e^{-\mathcal{L}t}\delta n_i P_{\rm eq}(n_i).
    \end{split}
     \label{eq:Paper2_TimeEvolOfni}
\end{equation}
Multiplying both sides of Eq.~\eqref{eq:Paper2_TimeEvolOfni} by $\delta n_i$ and then averaging, we obtain
\begin{equation}
   C(t)=\frac{\langle\delta n_i\rangle_t}{\langle \delta n_i\rangle_{t=0}},
   \label{eq:Paper2_GetCorrToNonequil}
\end{equation}
and hence Eq.~\eqref{eq:Paper2_CorrToNonEq}.

By separately setting the initial value of $n_i$ to $0$ and $1$, from Eq.~(\ref{eq:Paper2_GetCorrToNonequil}) we obtain two equivalent expressions for the correlation function:
\begin{equation}
    \begin{split}
        &C(t)=\frac{P[n_i(t)=1|n_i(0)=1]-c}{1-c},\\
        &C(t)=\frac{c-P[n_i(t)=1|n_i(0)=0]}{c}.
    \end{split}
    \label{eq:Paper2_CorrToReturnProb}
\end{equation}
These two probabilities in Eq.~\eqref{eq:Paper2_CorrToReturnProb} can be expressed by the path-probability densities:
\begin{equation}
    \begin{split}
        &P[n_i(t)=1|n_i(0)=1]=\\
        &\quad\sum_{m=0}^{\infty}\int_{0<t_1<\dots<t_{2m}<t}\Lambda_{2m}^1(\Delta {\bf t}_{2m+1})dt_1\dots dt_{2m},\\
        &P[n_i(t)=1|n_i(0)=0]=\\
        &\quad\sum_{m=1}^{\infty}\int_{0<t_1<\dots<t_{2m-1}<t}\Lambda_{2m-1}^0(\Delta {\bf t}_{2m})dt_1\dots dt_{2m-1}.
    \end{split}
    \label{eq:Paper2_ReturnProbToTrajectory}
\end{equation}
Taking the Laplace transform of Eq.~\eqref{eq:Paper2_ReturnProbToTrajectory}, replacing the path-probability densities with the decomposition in Eq.~\eqref{eq:Paper2_1stDecompOfLambdaXi} and substituting Eq.~\eqref{eq:Paper2_RelationOfPath&NextFlipDensity_nEqual0}, we obtain
\begin{equation}
    \begin{split}
        &\mathcal{L}\{P[n_i(t)=1|n_i(0)=1]\}=\\
        &\quad\quad\quad\quad\frac{z\tilde{\Lambda}_0^1(z)\tilde{\Xi}_0^0(z)+\tilde{\Xi}_0^1(z)-z\tilde{\Xi}_0^0(z)\tilde{\Xi}_0^1(z)}{z[\tilde{\Xi}_0^0(z)+\tilde{\Xi}_0^1(z)-z\tilde{\Xi}_0^0(z)\tilde{\Xi}_0^1(z)]},\\
        &\mathcal{L}\{P[n_i(t)=1|n_i(0)=0]\}=\\
        &\quad\quad\quad\quad\frac{[1-z\tilde{\Lambda}_0^0(z)]\tilde{\Xi}_0^1(z)}{z[\tilde{\Xi}_0^0(z)+\tilde{\Xi}_0^1(z)-z\tilde{\Xi}_0^0(z)\tilde{\Xi}_0^1(z)]}.
    \end{split}
    \label{eq:Paper2_ExpOfP11P10_SIF}
\end{equation}

Substituting Eq.~\eqref{eq:Paper2_ExpOfP11P10_SIF} into the Laplace transform of Eq.~\eqref{eq:Paper2_CorrToReturnProb}, we obtain two equivalent expressions for the correlation function $\tilde{C}(z)$ under 1-IF:
\begin{equation}
    \begin{split}
        &\tilde{C}^{\rm 1-IF}(z)=\\
        &\quad\frac{z\tilde{\Lambda}_0^1(z)\tilde{\Xi}_0^0(z)+\tilde{\Xi}_0^1(z)-z\tilde{\Xi}_0^0(z)\tilde{\Xi}_0^1(z)}{(1-c)z[\tilde{\Xi}_0^0(z)+\tilde{\Xi}_0^1(z)-z\tilde{\Xi}_0^0(z)\tilde{\Xi}_0^1(z)]}-\frac{c}{(1-c)z},\\
        &\tilde{C}^{\rm 1-IF}(z)=\\
        &\quad\frac{1}{z}-\frac{[1-z\tilde{\Lambda}_0^0(z)]\tilde{\Xi}_0^1(z)}{cz[\tilde{\Xi}_0^0(z)+\tilde{\Xi}_0^1(z)-z\tilde{\Xi}_0^0(z)\tilde{\Xi}_0^1(z)]}.
    \end{split}
    \label{eq:Paper2_CorrSIFTwoEq}
\end{equation}

\subsection{Three-point correlation function in terms of persistence functions}

The three-point correlation function in the East model is defined as follows:
\begin{equation}
    C_{1,2}(t)=\frac{\langle \delta n_i(t)\delta n_i(0)\delta n_{i+1}(0)\rangle}{c^{3/2}(1-c)^{3/2}}.
    \label{eq:Paper2_ThreeCorr}
\end{equation}
In the following, we relate it to the corresponding persistence functions.

Assume that the joint initial distribution of spin $i$ and spin $i+1$ is
\begin{equation}
    P_{t=0}(n_i,n_{i+1})=P_{\rm eq}(n_i)P_{\rm eq}(n_{i+1})\Delta_2(n_i,n_{i+1}),
    \label{eq:Paper2_initialDistriOfnini+1}
\end{equation}
where
\begin{equation}
    \Delta_2(n_i,n_{i+1})=1+a_1\delta n_i+a_2\delta n_{i+1}+f\delta n_i\delta n_{i+1}.
    \label{eq:Paper2_DeviationOfnini+1}
\end{equation}
The  parameters $a_1,a_2$ and $f$ can be expressed in terms of the initial averages of state variables:
\begin{equation}
    \begin{split}
        &a_1=\frac{\langle\delta n_i\rangle_{t=0}}{c(1-c)},\quad a_2=\frac{\langle\delta n_{i+1}\rangle_{t=0}}{c(1-c)},\quad f=\frac{\langle \delta n_i\delta n_{i+1}\rangle_{t=0}}{c^2(1-c)^2}.
    \end{split}
    \label{eq:Paper2_CoeffToInitialAve}
\end{equation}

Applying Eq.~\eqref{eq:Paper2_timeevolOfPnt} to Eq.~(\ref{eq:Paper2_initialDistriOfnini+1}) gives
\begin{equation}
    \begin{split}
        &P_t(n_i,n_{i+1})=\\
        &\quad\quad\Big(1+a_1e^{-\mathcal{L}t}\delta n_i+a_2e^{-\mathcal{L}t}\delta n_{i+1}\\
        &\quad\quad\quad\quad\quad\quad+fe^{-\mathcal{L}t}\delta n_i\delta n_{i+1}\Big)P_{\rm eq}(n_i)P_{\rm eq}(n_{i+1}).
    \end{split}
    \label{eq:Paper2_Paper2_TimeEvolOfnini+1}
\end{equation}
Multiplying both sides of Eq.~\eqref{eq:Paper2_Paper2_TimeEvolOfnini+1} by $\delta n_i$ and averaging, we obtain an expression for the three-point correlation function in terms of $\langle n_i\rangle_t$ and $C(t)$:
\begin{equation}
    C_{1,2}(t)=\sqrt{c(1-c)}\frac{\langle n_i\rangle_t-c-\langle \delta n_i\rangle_{t=0}C(t)}{\langle \delta n_i\delta n_{i+1}\rangle_{t=0}}.
    \label{eq:Paper2_ThreePointToNonEq}
\end{equation}
To connect $C_{1,2}(t)$ to the persistence functions, we set $n_{i+1}(t=0)=1$. We then obtain two expressions for $C_{1,2}(t)$, corresponding to $n_i(0)=1$ and $n_i(0)=0$, respectively:
\begin{equation}
    \begin{split}
        &C_{1,2}(t)=\sqrt{c(1-c)}\times\\
        &\frac{P[n_i(t)=1|n_i(0)=1,n_{i+1}(0)=1]-c-(1-c)C(t)}{(1-c)^2},\\
        &C_{1,2}(t)=\\
        &-\frac{P[n_i(t)=1|n_i(0)=0,n_{i+1}(0)=1]-c+cC(t)}{\sqrt{c(1-c)}}.
    \end{split}
    \label{eq:Paper2_C12tToReturnProb}
\end{equation}

These two probabilities in Eq.~\eqref{eq:Paper2_C12tToReturnProb} can be expressed in terms of the path-probability densities:
\begin{equation}
    \begin{split}
        &P[n_i(t)=1|n_i(0)=1,n_{i+1}=1]=\\
        &\quad\sum_{m=0}^{\infty}\int_{0<t_1<\dots<t_{2m}<t}\Xi_{2m}^1(\Delta {\bf t}_{2m+1})dt_1\dots dt_{2m},\\
        &P[n_i(t)=1|n_i(0)=0,n_{i+1}=1]=\\
        &\quad\sum_{m=1}^{\infty}\int_{0<t_1<\dots<t_{2m-1}<t}\Xi_{2m-1}^0(\Delta {\bf t}_{2m})dt_1\dots dt_{2m-1}.
    \end{split}
    \label{eq:Paper2_ReturnProbToTrajectoryC12}
\end{equation}
Taking the Laplace transform of Eq.~\eqref{eq:Paper2_ReturnProbToTrajectoryC12} and then applying the 1-IF to the functions $\tilde{\Xi}_{2m}^{1}(z{\bf 1}_{2m+1})$ and $\tilde{\Xi}_{2m-1}^{0}(z{\bf 1}_{2m})$, we obtain two  expressions of the three-point correlation function $\tilde{C}_{1,2}(z)$ under 1-IF:
\begin{align}
    \begin{split}
        &\frac{(1-c)^{3/2}}{\sqrt{c}}\tilde{C}_{1,2}(z)+(1-c)\tilde{C}(z)=\\
        &\quad\quad\quad\quad\frac{\tilde{\Xi}_0^1(z)}{z[\tilde{\Xi}_0^0(z)+\tilde{\Xi}_0^1(z)-z\tilde{\Xi}_0^0(z)\tilde{\Xi}_0^1(z)]}-\frac{c}{z},\\
        &\sqrt{c(1-c)}\tilde{C}_{1,2}(z)+c\tilde{C}(z)=\\
        &\quad\quad\quad\quad \frac{c}{z}-\frac{[1-z\tilde{\Xi}_0^0(z)]\tilde{\Xi}_0^1(z)}{z[\tilde{\Xi}_0^0(z)+\tilde{\Xi}_0^1(z)-z\tilde{\Xi}_0^0(z)\tilde{\Xi}_0^1(z)]}.
    \end{split}
    \label{eq:Paper2_C12SIFToPersist}
\end{align}

\subsection{Persistence functions in terms of correlation functions}
Combining Eqs.~\eqref{eq:Paper2_CorrSIFTwoEq} and \eqref{eq:Paper2_C12SIFToPersist} and solving these equations algebraically for the four persistence functions, we obtain
\begin{equation}
    \begin{split}
        &\tilde{\Lambda}_0^0(z)=\frac{(1-c)^{3/2}\tilde{C}_{1,2}(z)+\sqrt{c}\tilde{C}(z)}{(1-c)^{3/2}z\tilde{C}_{1,2}(z)+\sqrt{c}(1-c)z\tilde{C}(z)+c^{3/2}},\\
        &\tilde{\Lambda}_0^1(z)=\frac{\tilde{C}(z)+\sqrt{c(1-c)}\tilde{C}_{1,2}(z)}{1-c+\sqrt{c(1-c)}z\tilde{C}_{1,2}(z)+cz\tilde{C}(z)},\\
        &\tilde{\Xi}_0^0(z)=\frac{\sqrt{1-c}\tilde{C}_{1,2}(z)+\sqrt{c}\tilde{C}(z)}{(1-c)^{3/2}z\tilde{C}_{1,2}(z)+\sqrt{c}(1-c)z\tilde{C}(z)+c^{3/2}},\\
        &\tilde{\Xi}_0^1(z)=\frac{\sqrt{1-c}\tilde{C}_{1,2}(z)+\sqrt{c}\tilde{C}(z)}{\sqrt{c}-c^{3/2}+c\sqrt{1-c}z\tilde{C}_{1,2}(z)+c^{3/2}z\tilde{C}(z)}.
    \end{split}
    \label{eq:Paper2_PersistToCorrs}
\end{equation}

In summary, under the 1-IF approximation, the four persistence functions can be expressed in terms of the two correlation functions. On the other hand, the 2-DB approximation provides an explicit expression for the correlation function $\tilde{C}(z)$~\cite{van2005glassy}. 
Furthermore, the three-point correlation function is related to $\tilde{C}(z)$ via:
\begin{equation}
    \tilde{C}_{1,2}(z)=\frac{1-(z+c)\tilde{C}(z)}{\sqrt{c(1-c)}}.
    \label{eq:Paper2_C12ToC}
\end{equation}
We then directly compare the persistence functions calculated from these correlation functions via Eq.~\eqref{eq:Paper2_PersistToCorrs} with the 1-IF results, as shown in Fig.~\ref{fig:Paper2_PersistenceCompare}. 
Although a mathematical proof of their equivalence is still lacking, our numerical examination confirms that the two sets of theoretical results are nearly identical.

\bibliography{EastRef}
\end{document}